\documentclass[11pt]{JHEP3}

\usepackage{amsmath}
\usepackage{amssymb}
\usepackage{graphicx}
\usepackage{cite}

\newcommand{\be}{\begin{eqnarray}}
\newcommand{\ee}{\end{eqnarray}}
\newcommand{\nn}{\nonumber}
\newcommand{\bn}{\begin{enumerate}}
\newcommand{\en}{\end{enumerate}}

\def\Ds{D \hskip -7pt / \hskip 2pt}

\def\IC{\mathbb{C}}

\def\IZ{\mathbb{Z}}

\def\CL{{\cal L}}

\def\CN{{\cal N}}

\def\a{\alpha}
\def\b{\beta}
\def\g{\gamma}
\def\d{\delta}
\def\e{\epsilon}
\def\ve{\varepsilon}

\def\th{\theta}

\def\k{\kappa}
\def\l{\lambda}
\def\m{\mu}
\def\n{\nu}

\def\r{\rho}

\def\s{\sigma}

\def\t{\tau}

\def\w{\omega}
\def\G{\Gamma}
\def\D{\Delta}

\def\dd{\rm d}

\def\half{\frac{1}{2}}
\def\thalf{{\textstyle \frac{1}{2}}}

\def\goto{\rightarrow}

\def\da{{\dot{\a}}}
\def\db{{\dot{\b}}}
\def\dg{{\dot{\g}}}
\def\dd{{\dot{\d}}}
\def\tq{{\tilde{q}}}
\def\tps{{\tilde{\psi}}}

\newcommand{\ts}{\textstyle}

\newcommand{\sla}[1]{{/\!\!\!\!{#1}}}

\newcommand{\vev}[1]{\left\langle{#1}\right\rangle}

\newcommand{\ZZ}{{\mathbb Z}}

\title{ $\CN=4$ Superconformal Chern-Simons  Theories \\
with Hyper and Twisted Hyper Multiplets}

\author{Kazuo Hosomichi$^1$, Ki-Myeong Lee$^1$, Sangmin Lee$^2$, Sungjay Lee$^1$ and Jaemo Park$^{3,4,5}$

\\
\\
$^1$Korea Institute for Advanced Study, Seoul 130-012, Korea
\\
$^2$Department of Physics \& Astronomy, Seoul National University,
Seoul 151-747, Korea
\\
$^3$Department of Physics, POSTECH,
Pohang 790-784, Korea
\\
$^4$Postech Center for Theoretical Physics (PCTP), Postech, Pohang
  790-784, Korea
\\$^5$Department of Physics, Stanford University, Stanford, CA
94305-4060, USA
\\
\\
E-mail:
\email{hosomiti@kias.re.kr, klee@kias.re.kr, sangmin@snu.ac.kr, sjlee@kias.re.kr, jaemo@postech.ac.kr}
}

\abstract{ We extend the $\CN=4$ superconformal Chern-Simons
theories   of Gaiotto and Witten to those with additional twisted
hyper-multiplets. The new theories are generically linear quiver
gauge theories with the two types of hyper-multiplets alternating
between gauge groups. Our construction includes the Bagger-Lambert
model of  $SO(4)$ gauge group. A family of
abelian theories are identified with those proposed earlier in the
context of the M-crystal model
for M2-branes probing $(\IC^2/\IZ_n)^2$ orbifolds.
Possible extension with non-abelian BF
couplings and string/M-theory realization are briefly discussed. }

\preprint{KIAS-P08039 \\ SU-ITP-08/10}

\begin{document}


\section{Introduction}

Recently there has been considerable works on three-dimensional superconformal
theories with larger supersymmetries. Bagger and Lambert (BL) has
proposed an ${\cal N}=8$ superconformal model with three-algebra
structure and $SO(8)$ R-symmetry as a theory on multiple M2
branes \cite{BL1,BL2,BL3,gus1,gus2}. More recently Gaiotto and Witten (GW) has
proposed ${\cal N}=4$ superconformal Chern-Simons
models with $SO(4)$ R-symmetry classified by super-algebras \cite{gw}.

In this work we generalize the Gaiotto-Witten's work to include
 twisted hyper-multiplets.  Quiver theories appear naturally with
two types of hyper-multiplets alternating between gauge groups
where the quiver diagram is linear or circular with multiple nodes.
The Bagger-Lambert  theory with $SO(4)$ gauge group appears
naturally as a simplest kind of the quiver theory. Our work is partially
 motivated by attempt to understand the Bagger-Lambert  theory with
 $SO(4)$ gauge group in the context of the Gaiotto-Witten  theory.

The number of supersymmetries of three-dimensional superconformal Chern-Simons theories
has  a natural division with ${\cal N}=3$~\cite{Kao}. It is rather straightforward to have the
theories with ${\cal N}\le 3$, and there has been some recent work on ${\cal N}=2,3$
superconformal theories \cite{gy}. For the conformal theory of M2 branes, one
needs more supersymmetry~\cite{sch}
and the recent works related to the BL theory and the GW theory can be regarded as concrete steps toward this direction.

The BL theory was proposed as a superconformal field theory on 
coincident M2 branes. The BL theory is a superconformal
Chern-Simons theory with maximal supersymmetry and $SO(8)$
R-symmetry. It is parity even~\cite{sc2,rams} and its superconformal
symmetry group is $OSp(8|4)$.
The BL theory is supposed to describe two M2 branes
sitting at the origin of $\mathbb R^8$ with certain discrete
symmetry \cite{Lambert:2008et,Distler:2008mk}. For the additional many results on the BL theory, see
\cite{ho,ber1,muk,ber2,x0,x1,x2,x3,x4,HLL08,x5,x6,x7,x8,x9,bf1,bf2,bf3,y1,y2,y3,y4,y5,y6,y7,y8}.

Meanwhile, the GW theories are ${\cal N}=4$ superconformal
Chern-Simons theories with $SO(4)$ R-symmetry and $OSp(4|4)$
as the superconformal group. The GW theories have two gauge groups of
opposite Chern-Simons coefficient, and the possible gauge fields
are $U(N)\times U(M)$ or $O(N)\times Sp(M)$. The matter field
belongs to the bi-fundamental hyper-multiplet
$(q_\alpha^A,\psi_{\dot{\alpha}}^A)$ where the scalar field
$q^A_\alpha $ and the spinor field $\psi^A_{\dot{\alpha}}$ belong to
$({\bf 2},{\bf 1}) $ and $({\bf 1}, {\bf 2})$ representations of the
$SU(2)_L\times SU(2)_R=SO(4)$ R-symmetry group.

The GW theories were developed as  defect conformal field theories 
dual to the half-supersymmetric Janus geometry \cite{gut}. The four-dimensional Janus field theory
can have at most eight supersymmetries, suggesting that the
possible existence of the three-dimensional ${\cal N}=4$ superconformal field
theories. The GW theories form a complete class of the
superconformal Chern-Simons theories with a single bi-fundamental
hyper-multiplet.

On the other hand, the BL theory with $SO(4)$ gauge group has
$SO(8)$ R-symmetry. If one keeps only a half of the
supersymmetry, the R-symmetry group would be reduced to $SO(4)$, and the matter field splits to one hyper-multiplet and
one twisted hyper-multiplet. The twisted hyper-multiplet
$(\tilde{q}^A_{\dot{\alpha}}, \tilde{\psi}^A_\alpha)$ belongs to
$({\bf 1},{\bf 2}) $ and $({\bf 2}, {\bf 1})$ representations of the
$SU(2)_L\times SU(2)_R$ R-symmetry group. This suggests that
there could be a generalization of the GW theories with additional
twisted hyper-multiplet. Indeed, the present work realizes this
possibility.

Another motivation for our work arises from consideration of
the so-called M-crystal model~\cite{mc1, mc2, mc3}  of AdS$_4$/CFT$_3$ dual pairs.
This M-crystal model  is the M2-brane counterpart of
the famous brane-tiling model \cite{tile1,tile2,tile3} for
D3-brane gauge theories. The M-crystal model graphically encodes
certain information on the world-volume theories of M2-branes
probing toric Calabi-Yau four-fold (CY$_4$) cones. When the CY$_4$
cone is a product of two singular ALE spaces, the world-volume
theory has $\CN=4$ supersymmetry.

In Ref.~\cite{mc3}, it was shown  how to derive an abelian gauge theory
from a given M-crystal model. In particular, for
CY$_4=(\IC^2/\IZ_n)^2$, it was found that the abelian theory must
contain both types of hyper-multiplets, for a somewhat similar reason to the BL theory case.
 The derivation of
\cite{mc3} however was incomplete, as the kinetic terms for the abelian
gauge fields were not determined. Nevertheless, based on the analysis of
the moduli space of vacua, the gauge fields were
argued to be non-dynamical. Naturally, our work provides
a possible candidate theory for the corresponding CFT$_3$.

In section 2, we begin by reviewing the GW construction.
In this construction, one uses the ${\cal N}=1$ theory with an
$SO(3)$ global symmetry
and adjusts the coupling constants to enhance the global symmetry to the  $SO(4)$ R-symmetry,
which does not commute with the ${\cal N}=1$ supercharge. The resulting theory has ${\cal N}=4$ supersymmetry. We employ a similar method to include additional twisted
hyper-multiplet to the theory and find new ${\cal N}=4$ superconformal field theories.

As in Ref.~\cite{gw}, the gauge groups and the matter contents are
classified by certain super-algebras. The restriction is quite
severe, so that to have non-trivial interactions between the two
types of hyper-multiplets, the theory must be a linear quiver
gauge theory (possibly forming a closed loop) with the two types
of hyper-multiplets alternating between neighboring gauge groups.
We also present how to take a mass deformation of the
general $\CN=4$ theories without breaking any supersymmetry or $SO(4)$ R-symmetry.

In section 3, we discuss how our construction is related to other
works. First, we show explicitly that the  BL model of $SO(4)$ gauge group with its mass
deformation~\cite{x3,HLL08}  is a special case
of our construction. For this case, the supersymmetry is doubled by chance. However, we suspect that
our construction for two types of hyper-multiplets gets an enhanced supersymmetry only for the
BL case.   The second half is devoted to the connection to the M-crystal model. After a short review of the M-crystal model specialized to $\CN=4$, we show that a particular abelian quiver theory of the present paper can be identified with the one proposed in \cite{mc3} for M2-branes probing $(\IC^2/\IZ_n)^2$ orbifolds.

We conclude with some discussions on future directions in section 4.


\section{  $\mathbf{\CN=4}$ Chern-Simons  Theories with Two Types of Hyper-Multiplets}

In this section, we start with a brief review of the GW construction
of the ${\cal N}=4$
superconformal theories with only hyper-multiplets, and then present a generalization to include twisted hyper-multiplets.

\subsection{Gaiotto-Witten revisited}

We start with an $Sp(2n)$ group and let $A,B$ indices run over a $2n$-dimensional representation.
We denote the anti-symmetric invariant tensor of $Sp(2n)$ by $\omega_{AB}$ and choose all the generators $t^A_{~B}$ to be anti-Hermitian $(2n\times 2n)$ matrices, such that $(\omega_{AC}t^C_{~B})$ are symmetric matrices. We consider a  Chern-Simons gauge theory whose gauge group is a subgroup of $Sp(2n)$ and we denote  anti-Hermitian generators of the gauge group as $(t^m)^A_{~B}$ which satisfy
\[
 [t^m,t^n]=f^{mn}{}_p t^p.
\]
Gauge field and gaugino are denoted by $(A_m)_\mu$ and $\chi_m$, and
the adjoint indices are raised or lowered by an invariant quadratic
form $k^{mn}$ or its inverse $k_{mn}$ of the gauge group.
We will also use
\[
 \chi^A_{~B}=\chi_m (t^m)^A_{~B},~~~~
 \chi_{AB}= \chi_m \omega_{AC} (t^m)^C{}_B = \chi_m t^m_{AB}.
\]

We couple the gauge theory with a
hyper-multiplet matter fields  $(q^A_\a, \psi^A_\da)$ satisfying the reality condition
\[
 \bar q_A^\a=(q^A_\a)^\dagger=\epsilon^{\a\b}\omega_{AB}q^B_\b,~~~~~~
 \bar \psi_\da^A=(\psi^A_\da)^\dagger=\epsilon^{\da\db}\omega_{AB}\psi^B_\db.
\]
We use $(\a,\b ; \da,\db)$ doublet indices for the $SU(2)_L\times SU(2)_R$
R-symmetry group.

To get $\CN=4$ supersymmetric theories, it is convenient to work
in the $\CN=1$ framework in which
$(q^A_\alpha,\psi^A_{\dot\alpha},F^A_\alpha)$ and $((A_m)_\mu,\chi_m)$ are the
basic supermultiplets.
Our conventions for spinors and $\CN=1$ superfields are
summarized in appendix A. We start from the general $\CN=1$
Lagrangian with a global $SU(2)$ symmetry which acts on the
indices $(\alpha,\beta)$ and $(\dot\alpha,\dot\beta)$
simultaneously. Such a Lagrangian would take  the form ${\cal L} = {\cal
L}_{\rm CS}+{\cal L}_{\rm kin}+{\cal L}_W$, where
\begin{eqnarray}
 {\cal L}_{\rm CS} &=&
  \frac{\varepsilon^{\m\n\l}}{4\pi}k_{mn}A_\m^m\partial_\n A_\l^n
  +\frac{\varepsilon^{\m\n\l}}{12\pi}f_{mnp} A_\m^m A_\n^n A_\l^p
  +\frac{ik_{mn}}{4\pi}\chi^m\chi^n,
\nn\\ %
 {\cal L}_{\rm kin} &=&
  \frac12
  \left(
   -D \bar q_A^\a D q^A_\a
   +\bar F_A^\a F^A_\a
   +i\bar\psi_A^\da \Ds \psi^A_\da
   -i\bar\psi_A^\da \chi^A_{~B} q^B_\a
   +i\bar q_A^\a \chi^A_{~B} \psi^B_\da
  \right)
\nn\\ %
  &=&  \frac12\omega_{AB}
  \left[
   \epsilon^{\a\b}\left(-D q^A_\a D q^B_\b + F^A_\a F^B_\b\right)
   +i\epsilon^{\da\db} \psi^A_\da \Ds \psi^B_\db
  \right]
 -i\epsilon^{\da\b}
   \psi^A_\da\chi_{AB}q^B_\b,
\nn\\ %
 {\cal L}_W
  &=& - \pi T_{AB,CD}
 \left(
  \epsilon^{\a\b}\epsilon^{\g\d} F_\a^A q_\b^B q_\g^C q_\d^D
 +\frac{i}{2} \epsilon^{\da\db}\epsilon^{\g\d} \psi_\da^A \psi_\db^B q_\g^C q_\d^D
 +i\epsilon^{\da\b}\epsilon^{\dg\d} \psi_\da^A q_\b^B \psi_\dg^C q_\d^D
 \right), \;\;\;\;
 \end{eqnarray}
 where $D_\mu q^A_\alpha =\partial_\mu q^A_\alpha + A_{m\mu} (t^m)^A_{~B}q^B_\alpha$.
 The superpotential takes the form
\begin{equation}
 W  = \frac{\pi}{4}T_{AB,CD}
 \epsilon^{\a\b}\epsilon^{\g\d}q_\a^A q_\b^B q_\g^C q_\d^D.
\end{equation}
Here the superpotential coupling $T_{AB,CD}$ is anti-symmetric under
the permutation of $A,B$ or $C,D$, and symmetric under the exchange of the
pairs $(AB)$ with $(CD)$. For a suitable choice of $T_{AB,CD}$, the Lagrangian becomes invariant under
two $SU(2)$ rotations that act on $q_\alpha^A$ and $\psi_{\dot\alpha}^A$
separately and therefore $\CN=4$ supersymmetric.
Let us now examine each part of the Lagrangian which needs to be
separately invariant under the $SO(4)$.

\paragraph{Yukawa coupling}

As the gaugino field $\chi_m$ is a purely auxiliary field, we integrate it out to get a $(q^2\psi^2)$ term,
\begin{eqnarray}
{\cal L} &=&
 \frac{ik_{mn}}{4\pi}\chi^m\chi^n
 -i\chi_m \left(\epsilon^{\da\b}\psi_\a^A t^m_{AB}q^B_\b\right) + \cdots
 \nn\\ &=& \frac{ik_{mn}}{4\pi}
  \left(\chi^m-2\pi\epsilon^{\da\b}
        \psi^A_\da t^m_{AB}q^B_\b\right)
  \left(\chi^n-2\pi\epsilon^{\dg\d}
        \psi^C_\dg t^n_{CD}q^D_\d\right)
  \nn\\&&
 -i\pi k_{mn} t^m_{AB} t^n_{CD}
   \epsilon^{\da\b}\epsilon^{\dg\d}\psi^A_\da q^B_\b \psi^C_\dg q^D_\d
 +  \cdots.
\end{eqnarray}
Combining this with the other $(q^2\psi^2)$ terms arising from
the superpotential ${\cal L}_W$, one finds
\begin{equation}
 -i\pi q_\a^A q_\b^B \psi_\dg^C \psi_\dd^D
 \left(
   k_{mn} t^m_{AC} t^n_{BD} \epsilon^{\a\dg}\epsilon^{\b\dd}
   + \thalf T_{AB,CD}\epsilon^{\a\b}\epsilon^{\dg\dd}
   + T_{AC,BD}\epsilon^{\a\dg}\epsilon^{\b\dd}
 \right).
\end{equation}
This expression has to be $SU(2)_L\times SU(2)_R$-invariant on its own.
It implies that the terms containing contractions
between dotted and undotted indices must vanish.
For the Yukawa coupling, it suffices to require that
the part proportional to $q_{(\a}^{(A}q_{\b)}^{B)}$ should vanish:
\begin{equation}
  T_{AC,BD}
+ T_{BC,AD}
+ k_{mn} \left( t^m_{AC} t^n_{BD}
+ t^m_{BC} t^n_{AD} \right) ~=~ 0.
\label{TTtt}
\end{equation}
In Ref.~\cite{gw}, this equation was shown to determine the coupling constants, 
\begin{equation}
 T_{AB,CD}~=~ \frac13 k_{mn} \left(
 t^m_{AC} t^n_{BD}- t^m_{BC} t^n_{AD}
 \right).
\end{equation}
and imposes a constraint on $t^m_{AB}$ which the authors called
the ``fundamental identity'', 
\be
\label{fundi}
k_{mn} t^m_{(AB}t^n_{C)D}=0 ,
\ee
where the indices $A,B,C$ are symmetrized over cyclic permutations.
In Ref.~\cite{gw}, it was also noticed that this identity can be
understood as the Jacobi identity for three fermionic generators
of a super Lie algebra,
\begin{equation}
  [M^m,M^n]=f^{mn}_{~~~p}M^p,~~~~
  [M^m,Q_A]=Q_B (t^m)^B_{~A},~~~~
  \{Q_A,Q_B\}= t^m_{AB} M_m.
\end{equation}
This turns out to be a rather strong constraint on the field content
of the theory.
Namely, the gauge group and matter should be such that the gauge
symmetry algebra can be extended to a super Lie algebra by adding
fermionic generators associated to hyper-multiplets.
The final expression for the $(q^2\psi^2)$ term in the Lagrangian is
\be
 {\cal L}_{q^2\psi^2} &=& -i\pi q_\a^A q_\b^B \psi_\g^C \psi_\d^D \epsilon^{\a\b}\epsilon^{\dg\dd} k_{mn} t^m_{AC} t^n_{BD}
\nn \\
&=& -i\pi k_{mn} \e^{\a\b} \e^{\dg\dd} \jmath^m_{\a\dg} \jmath^n_{\b\dd}
\;\;\;\;\;
\left( \jmath^m_{\a\dg} \equiv q_\a^A t^m_{AC} \psi^C_\dg \right).
\ee

\paragraph{Bosonic potential}
We present here the computation of bosonic potential in some
detail for later convenience.
In terms of the ``moment map'', $\mu^m_{\a\b}\equiv t^m_{AB}q^A_\a q^B_\b$ \cite{gw},
we can write
\begin{equation}
 W =  \frac\pi6\epsilon^{\a\b}\epsilon^{\g\d}k_{mn}\mu^m_{\a\g}\mu^n_{\b\d} .
\end{equation}
To compute the bosonic potential, it is useful to note
\be
 \omega^{AB}\epsilon_{\a\b}
 \frac{\partial\mu_{\g\d}^m}{\partial q^A_\a}
 \frac{\partial\mu_{\k\r}^n}{\partial q^B_\b} &=&
 \epsilon_{\g\k}\mu^{mn}_{\d\r}+\epsilon_{\g\r}\mu^{mn}_{\d\k}+
 \epsilon_{\d\k}\mu^{mn}_{\g\r}+\epsilon_{\d\r}\mu^{mn}_{\g\k},
\ee
where
\be
 \mu^{mn}_{\a\b} &\equiv& (\w t^m t^n)_{AB} q^A_\a q^B_\b.
\nn
\ee
The potential term is
\begin{eqnarray}
 V &=&
 \frac12\epsilon_{\a\b}\w^{AB}
 \frac{\partial W}{\partial q_\a^A}\frac{\partial W}{\partial q_\b^B}
 ~=~
 \frac{2\pi^2}{9}\epsilon_{\a\g}\mu^{mn}_{\b\d}\mu_m^{\a\b}\mu_{n}^{\g\d}
 \nn\\ &=&
 \frac{\pi^2}{9}\epsilon_{\a\g}f^{mnp}\mu_{p,\b\d}\mu_m^{\a\b}\mu_{n}^{\g\d}
+\frac{\pi^2}{9}\epsilon_{\a\g}\epsilon_{\b\d}\epsilon_{\k\r}
 \mu^{mn,\k\r} \mu_m^{\a\b}\mu_n^{\g\d}.
\label{Vq6}
\end{eqnarray}
As explained in Ref.~\cite{gw}, we apply the fundamental identity (\ref{fundi}) to rotate the upper indices $(\a\b\k)$
in the second term.
After some manipulations, we find
\begin{equation}
 V ~=~
 \frac{\pi^2}{9}\epsilon_{\a\g}f^{mnp}\mu_{p,\b\d}\mu_m^{\a\b}\mu_{n}^{\g\d}
-\frac{2\pi^2}{9}\epsilon_{\a\g}\mu_{\b\d}^{nm}\mu_m^{\a\b}\mu_{n}^{\g\d}.
\label{Vq62}
\end{equation}
Taking the average of the first line of (\ref{Vq6}) and (\ref{Vq62}),
we arrive at the final expression,
\begin{equation}
 V~=~\frac{\pi^2}{6} f_{mnp} (\mu^m)^\a_{~\b}(\mu^n)^\b_{~\g}(\mu^p)^\g_{~\a}.
\end{equation}

\paragraph{Full theory}

To summarize, the Gaiotto-Witten theory in a
manifestly ${\cal N}=4$ supersymmetric notation without auxiliary fields
consists of the Lagrangian
\begin{eqnarray}
 {\cal L} &=&
 \frac{\varepsilon^{\m\n\l}}{4\pi}
 \left(k_{mn} A^m_\m \partial_\n A^n_\l
      +\frac13 f_{mnp} A^m_\m A^n_\n A^p_\l \right)
 +\frac12\omega_{AB}
  \left(-\epsilon^{\a\b} D q^A_\a D q^B_\b+i\epsilon^{\da\db}\psi^A_\da \Ds\psi^B_\db \right)
 \nn\\&&
 -i\pi k_{mn} \e^{\a\b} \e^{\dg\dd} \jmath^m_{\a\dg} \jmath^n_{\b\dd}  - \frac{\pi^2}{6} f_{mnp} (\mu^m)^\a_{~\b}(\mu^n)^\b_{~\g}(\mu^p)^\g_{~\a},
\label{Lful1}
\end{eqnarray}
and the supersymmetry transformation rules
\be
 &&\delta q_\a^A = i\eta_\a{}^{\da} \psi_\da^A,~~~~
 \delta A^m_\m = 2\pi i \eta^{\a\da} \gamma_\m \jmath^m_{\a\da},
  \nn \\
 &&\delta\psi_\da^A=
 \left[\Ds q_\a^A + \frac{2\pi}3(t_m)^A_{~B} q^B_\b (\mu^m)^\b_{~\a}\right]
 \eta^\a_{~\;\da}.
\ee
The supersymmetry parameter $\eta$ is a Majorana spinor
and transform in the $(\mathbf{2,2})$ representation of $SU(2)_L \times SU(2)_R$.

The classical supergroups related to the GW models are $U(N|M)$ and $OSp(N|M)$. The gauge groups
are product groups $U(N)\times U(M)$ or $O(N)\times Sp(M)$ with equal and opposite Chern-Simons coefficients
for the two factors. The matter fields belong to a bi-fundamental representation of the product gauge group.
Of course, one can have a multiple embedding  of $t^m$ in $Sp(2n)$, resulting in many copies of the GW models, possibly with different gauge group pairs, and no coupling between different blocks of GW models.

\subsection{Adding twisted hyper-multiplets}

Let us try to include the twisted hyper-multiplets.
We denote them by $(\tilde{q}^A_\da, \tilde{\psi}^A_\a, \tilde{F}^A_\da$)
and define their moment map and its super-partner \cite{gw} as in
the untwisted case.
\be
\tilde{\m}^m_{\da\db} \equiv \tilde{t}^m_{AB} \tilde{q}^A_\da \tilde{q}^B_\db
, ~~~~
\tilde{\jmath}^m_{\da \a} \equiv \tilde{q}^A_\da \tilde{t}^m_{AB} \tilde{\psi}^B_\a .
\ee
Both types of hyper-multiplets share the same gauge symmetry,
so the structure constants $f^{mn}_{~~~p}$ and the quadratic
form $k^{mn}$ are identical.
However, they can take different representations,
so in general the generators $\tilde t^m$ are different from $t^m$.
Strictly speaking, we should distinguish the $A,B$ indices
of hyper-multiplets from those of twisted hyper-multiplets,
but we suppress the distinction to avoid clutter.

The construction of ${\cal L}_{\rm CS}$ and ${\cal L}_{\rm kin}$
proceeds in the same way as before.
The Yukawa and bosonic potential terms become more complicated due
to the mixing of the hyper and twisted hyper-multiplets.
We need to make those terms $SU(2)_L\times SU(2)_R$ R-symmetric
by a suitable choice of the superpotential.
The unmixed terms can all be made R-symmetric by introducing the super-potential
\begin{equation}
 W_0 =\frac\pi6\epsilon^{\a\b}\epsilon^{\g\d}k_{mn}\mu^m_{\a\g}\mu^n_{\b\d}
  +\frac\pi6\epsilon^{\da\db}\epsilon^{\dg\dd}k_{mn}\tilde\mu^m_{\da\dg}\tilde\mu^n_{\db\dd},
\end{equation}
provided that the fundamental identity (\ref{fundi}) holds
for both of the matrices $t^m_{AB}$ and $\tilde t^m_{AB}$.
The hyper and twisted hyper-multiplets therefore give two
(in general independent) extensions of the gauge symmetry algebra
to a super Lie algebra.

There may be many copies of the simple GW models in many pairs of
the gauge groups related to the supergroups. The key point is that the gauge group
pairs for the twisted hyper-multiplets do not need to coincide with those for
the hyper-multiplets. This allows the interaction between different copies of the GW model
for hyper-multiplets, leading to a quiver theory. While there may be many sets of quivers, one
can focus on one irreducible quiver theory where every part of theory are interacting with each other.

Now we focus on possible R-symmetry breaking terms due to the mixed interactions.

\paragraph{Yukawa coupling}
We first integrate out the gaugino,
\begin{equation}
 \chi^m=
 2\pi\left( \epsilon^{\da\b}\psi_\da^A t^m_{AB}q^B_\b
+ \epsilon^{\a\db}\tilde{\psi}_\a^A
 \tilde{t}^m_{AB}\tilde{q}^B_\db \right).
\end{equation}
Computing the Yukawa term, we find
a cross term from the gaugino squared,
\begin{equation}
 {\cal L}_{\rm Yukawa} ~=~
 -2\pi i\epsilon^{\a\db}\epsilon^{\dg\d}k_{mn}t^m_{AB}
   \tilde{t}^n_{CD}
   q^A_\a\psi^B_\db \tilde{q}^C_\dg \tilde{\psi}^D_\d + ({\rm unmixed}),
\label{GW_y_cr}
\end{equation}
which is not $SO(4)_R$ invariant by itself.
To restore the R-symmetry, we try adding some mixed terms to
the superpotential.
The most general form allowed by the diagonal $SU(2)$ and gauge
symmetries is
\begin{equation}
 \Delta W ~=~
 \pi\tilde{S}_{AB,CD}
 \epsilon^{\a\b}\epsilon^{\dg\dd} q_\a^A q_\b^B \tq_\dg^C \tq_\dd^D
+\pi S_{AB,CD}
 \epsilon^{\a\dg}\epsilon^{\b\dd} q_\a^A q_\b^B \tq_\dg^C \tq_\dd^D,
\end{equation}
where the coupling constants $\tilde{S}$ and $S$ satisfy
\[
  \tilde{S}_{AB,CD}=
 -\tilde{S}_{BA,CD}=
 -\tilde{S}_{AB,DC},~~~~~~
  S_{AB,CD}=
  S_{BA,CD}=
  S_{AB,DC}.
\]
The additional superpotential yields some Yukawa terms
which are themselves R-invariant,
\begin{eqnarray}
\Delta{\cal L}^{(1)}_{\rm Yukawa} &=&
-i\pi\tilde{S}_{AB,CD}\left( \epsilon^{\da\db}\epsilon^{\dg\dd}
  \psi_\da^A\psi_\db^B \tq_\dg^C \tq_\dd^D
  + \epsilon^{\a\b}\epsilon^{\g\d} q_\a^A q_\b^B \tps_\g^C \tps_\d^D \right)
\\&&
 -i\pi S_{AB,CD}\left( \epsilon^{\da\dg}\epsilon^{\db\dd}
 \psi_\da^A\psi_\db^B \tq_\dg^C \tq_\dd^D
 +\epsilon^{\a\g}\epsilon^{\b\d}
   q_\a^A q_\b^B \tps_\g^C \tps_\d^D
 +2\epsilon^{\a\g}\epsilon^{\db\dd}
 q_\a^A \psi_\db^B \tps_\g^C \tq_\dd^D \right), \nn
\end{eqnarray}
and those which are not,
\begin{eqnarray}
\Delta{\cal L}^{(2)}_{\rm Yukawa} &=&
 -2\pi i\left(
       2\tilde{S}_{AB,CD}\epsilon^{\a\db}\epsilon^{\dg \d}
       +S_{AB,CD}\epsilon^{\a\dg}\epsilon^{\db\d}
 \right)
  q_\a^A \psi_\db^B \tq_\dg^C \tps_\d^D .
\label{GW_y_cr2}
\end{eqnarray}
It is possible to combine (\ref{GW_y_cr}) and (\ref{GW_y_cr2})
in an R-symmetric way,
\be
 k_{mn} t^m_{AB}\tilde t^n_{CD}
  \epsilon^{\a\db}\epsilon^{\dg \d}
 +2\tilde{S}_{AB,CD}\epsilon^{\a\db}\epsilon^{\dg \d}
       +S_{AB,CD}\epsilon^{\a\dg}\epsilon^{\db\d}
       ~\sim~ \epsilon^{\a\d}\epsilon^{\db\dg},
\ee
by using an identity for the diagonal $SU(2)$:
$
\e^{\a\db}\e^{\dg \d} + \e^{\a\dg}\e^{\d \db} + \e^{\a\d}\e^{\db \dg} =0.
$
That uniquely determines the two coupling constants of $\D W$:
\begin{equation}
  \tilde{S}_{AB,CD}=0,~~~~
  S_{AB,CD}=- k_{mn} t^m_{AB}\tilde t^n_{CD}.
\end{equation}

\paragraph{Bosonic potential}
The superpotential is $W=W_0+\Delta W$, where
\begin{eqnarray}
\label{w0}
  W_0 &=& \frac\pi6\epsilon^{\a\b}\epsilon^{\g\d}k_{mn}\mu^m_{\a\g}\mu^n_{\b\d}
  +\frac\pi6\epsilon^{\da\db}\epsilon^{\dg\dd}k_{mn}\tilde\mu^m_{\da\dg} \tilde\mu^n_{\db\dd},
 \\
 \Delta W &=& -\pi\epsilon^{\a\db}\epsilon^{\g\dd}k_{mn}\mu^m_{\a\g}\tilde\mu^n_{\db\dd}.
 \label{w1}
\end{eqnarray}
We need to compute the mixed term $V|_{q^4\tilde q^2}$ in the bosonic
potential and check its R-invariance.
The computation of the other mixed term $V|_{q^2\tilde q^4}$ is similar.
There are two contributions to $V|_{q^4\tilde q^2}$,
\begin{eqnarray}
 V|_{q^4\tilde q^2} &=&
 \epsilon_{\a\b}\w^{AB}
 \frac{\partial W_0}{\partial q_\a^A}\frac{\partial\Delta W}{\partial q_\b^B}
+\frac12\epsilon_{\da\db}\w^{AB}
 \frac{\partial\Delta W}{\partial\tilde q_\da^A}
 \frac{\partial\Delta W}{\partial\tilde q_\db^B}
 \nn\\ &=&
 -\frac{4\pi^2}{3}\epsilon_{\a\dg}\mu^{mn}_{\b\d}\mu_m^{\a\b}\tilde\mu_{n}^{\dg\dd}
 +2\pi^2\epsilon_{\a\g}\tilde\mu^{mn}_{\db\dd}\mu_m^{\a\b}\mu_{n}^{\g\d}.
\end{eqnarray}
Using the trick explained below (\ref{Vq6}) again,
we can rewrite the first term as
\[
 -\pi^2f_{mnp}(\mu^m)^\a_{~\b}(\mu^n)^\b_{~\g}(\tilde\mu^p)^\dg_{~\da}.
\]
The second term is decomposed into those proportional to
$\tilde\mu^{[mn]}_{(\db\dd)}$ or $\tilde\mu^{(mn)}_{[\db \dd]}$.
The former precisely cancels against the above (first) term,
and we are left with the latter which is indeed R-invariant,
\begin{eqnarray}
 V|_{q^4\tilde q^2} &=&-\pi^2(\tilde\mu^{mn})^\dg_{~\dg}(\mu_m)^\a_{~\b}(\mu_n)^\b_{~\a}.
\end{eqnarray}
Combining all the mixed and unmixed terms, we obtain the full bosonic potential,
\begin{eqnarray}
 V &=&
   \frac{\pi^2}6f_{mnp}(\mu^m)^\a_{~\b}(\mu^n)^\b_{~\g}(\mu^p)^\g_{~\a}
  +\frac{\pi^2}6f_{mnp}
   (\tilde\mu^m)^\a_{~\b}(\tilde\mu^n)^\b_{~\g}(\tilde\mu^p)^\g_{~\a}
 \nn\\&&
  -\pi^2(\tilde\mu^{mn})^\dg_{~\dg}(\mu_m)^\a_{~\b}(\mu_n)^\b_{~\a}
  -\pi^2(\mu^{mn})^\g_{~\g}(\tilde\mu_m)^\da_{~\db}(\tilde\mu_n)^\db_{~\da}.
\end{eqnarray}

\paragraph{Full theory}

In summary, we have found the generalization of Gaiotto-Witten theory
which includes both hyper and twisted hyper-multiplets.
The full Lagrangian is given by
\begin{eqnarray}
 {\cal L} &=&
 \frac{\varepsilon^{\m\n\l}}{4\pi}
 \left(k_{mn} A^m_m \partial_\n A^n_\l
      +\frac13 f_{mnp} A^m_\m A^n_\n A^p_\l \right)
\nn\\&&
 +\frac12\omega_{AB}
  \left(-\epsilon^{\a\b} D q^A_\a D q^B_\b
        +i\epsilon^{\da\db}\psi^A_\da \Ds\psi^B_\db \right)
 +\frac12\tilde\omega_{AB}
  \left(-\epsilon^{\da\db} D \tilde q^A_\da D \tilde q^B_\db
       +i\epsilon^{\a\b}\tilde\psi^A_\a \Ds\tilde\psi^B_\b \right)
 \nn\\&&
 -i\pi k_{mn} \e^{\a\b} \e^{\dg\dd}\jmath^m_{\a\dg}\jmath^n_{\b\dd}
 -i\pi k_{mn} \e^{\da\db} \e^{\g\d}\tilde\jmath^m_{\da\g}\tilde\jmath^n_{\db\d}
 +4\pi ik_{mn}\epsilon^{\a\g}\epsilon^{\db\dd}
   \jmath^m_{\a\db}\tilde\jmath^n_{\dd\g}
 \nn\\&&
 +i\pi k_{mn}\left(
 \epsilon^{\da\dg}\epsilon^{\db\dd}
 \tilde\mu^m_{\da\db} \psi_\dg^At^n_{AB}\psi_\dd^B
 +\epsilon^{\a\g}\epsilon^{\b\d}
  \mu^m_{\a\b}\tps_\g^A\tilde t^n_{AB}\tps_\d^B
 \right)
 \nn\\&&
  -\frac{\pi^2}6f_{mnp}(\mu^m)^\a_{~\b}(\mu^n)^\b_{~\g}(\mu^p)^\g_{~\a}
  -\frac{\pi^2}6f_{mnp}
   (\tilde\mu^m)^\a_{~\b}(\tilde\mu^n)^\b_{~\g}(\tilde\mu^p)^\g_{~\a}
 \nn\\&&
  +\pi^2(\tilde\mu^{mn})^\dg_{~\dg}(\mu_m)^\a_{~\b}(\mu_n)^\b_{~\a}
  +\pi^2(\mu^{mn})^\g_{~\g}(\tilde\mu_m)^\da_{~\db}(\tilde\mu_n)^\db_{~\da},
\label{Lful2}
\end{eqnarray}
and the supersymmetry transformation law is
\begin{eqnarray}
 && \delta q_\alpha^A
    = +i\eta_\alpha^{~\;\dot\alpha}\psi_{\dot\alpha}^A,
~~~
    \delta\tilde q_{\dot\alpha}^A
    = -i\eta_{~\;\dot\alpha}^{ \alpha}\tilde\psi_\alpha^A,
~~~
    \delta A^m_\mu = 2\pi i\eta^{\alpha\dot\alpha}\gamma_\mu
            (\jmath^m_{\alpha\dot\alpha}-\tilde\jmath^m_{\dot\alpha\alpha}),
 \nn\\
 && \delta\psi_{\dot\alpha}^A
    = +\left[\sla Dq_\alpha^A
           +\frac{2\pi}3(t_m)^A_{~B}q^B_\beta(\mu^m)^\beta_{~\alpha}\right]
            \eta^\alpha_{~\;\dot\alpha}
           -2\pi(t_m)^A_{~B}q^B_\beta(\tilde\mu^m)^{\dot\beta}_{~\dot\alpha}
            \eta^\beta_{~\;\dot\beta} ,
 \nn\\
 && \delta\tilde\psi_\alpha^A
    = -\left[\sla D\tilde q_{\dot\alpha}^A
            +\frac{2\pi}3(\tilde t_m)^A_{~B}\tilde q^B_{\dot\beta}
             (\tilde\mu^m)^{\dot\beta}_{~\dot\alpha}\right]
            \eta_\alpha^{~\;\dot\alpha}
           +2\pi(\tilde t_m)^A_{~B}\tilde q^B_{\dot\beta}
             (\mu^m)^{\beta}_{~\alpha}
            \eta_\beta^{~\;\dot\beta} .
            \label{Sful2}
\end{eqnarray}

\paragraph{Classification in terms of quivers}

\begin{figure}[t]
\begin{center}
\includegraphics[width=9cm]{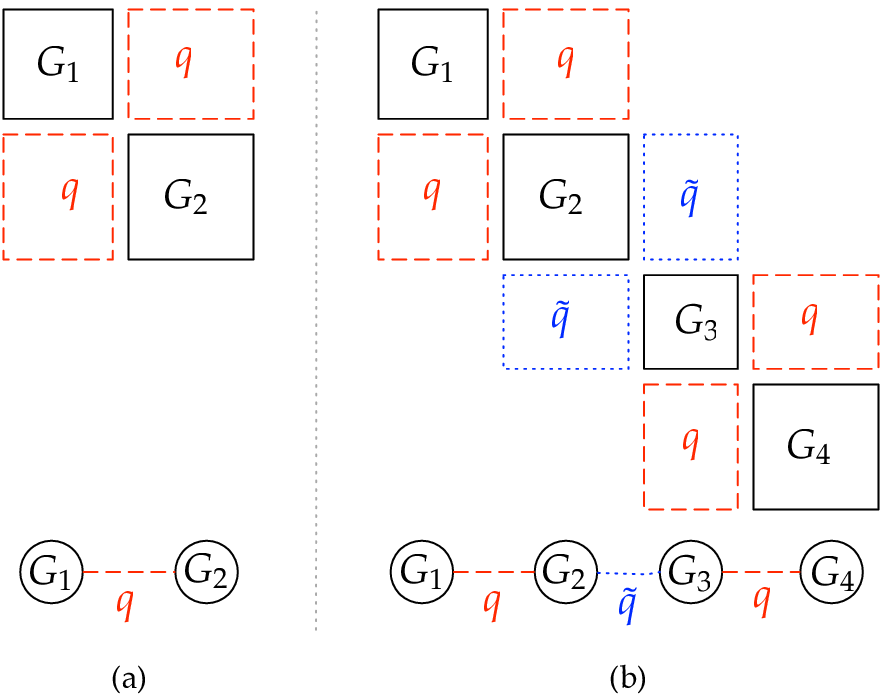}
\caption{Linear quiver structure of the original and extended GW theories. The gauge groups can be either $U(N_i)$ or alternation between $O(N_i)$ and $Sp(M_i)$. Dashed lines denote hyper-multiplets and dotted lines
denote twisted hyper-multiplets.
} \label{quiver}
\end{center}
\end{figure}
When only hyper-multiplets are considered, the
classification of $\CN=4$ Chern-Simons-matter theories boils down
to that of super Lie algebras.
In the purely non-abelian case,
the only possibilities are either the basic models $U(N|M)$ and $OSp(N|M)$
or multiple copies of them; see Figure \ref{quiver}(a).
Naive attempts to obtain more general quiver theories by
connecting several gauge groups with bi-fundamental hyper-multiplets
immediately ruin the super Lie algebra structure.

Once both hyper and twisted hyper-multiplets are included, we have
a richer variety of theories characterized by quiver diagrams
depicted in Figure \ref{quiver}(b).
If we align the gauge group factors linearly
and introduce bi-fundamental matter fields such that
the two types of hyper-multiplets alternate among
the gauge groups, we can have an interacting theory
of both types of hyper-multiplets
without violating the fundamental identity (\ref{fundi}).
Note that the hyper and twisted hyper-multiplets give two different ways
to pair the gauge groups into supergroups.
We should emphasize that, apart from trivial direct sums,
these linear quiver theories exhaust all possible (purely non-abelian) theories.

The linear quiver can either have open ends or form a closed circular loop.
With $U(N_i)$ gauge groups, the relevant quiver diagrams are the Dynkin diagrams for $A_n$ or $\hat{A}_{2n-1}$, respectively.
The BL-like models, which have only two gauge group factors
and both hyper-multiplets in the same representation, can be thought of as the shortest closed loops.


\paragraph{Abelian CS/BF theories}

Abelian theories deserve separate treatment.
The theories are defined by $U(1)$ gauge fields $A_m$,
quaternion-valued (twisted) hyper-multiplets $(q^i, \tilde{q}^{\bar{\imath}})$,
the charge matrices $(Q^m_i, \widetilde{Q}^m_{\bar{\imath}})$,
and the CS coefficient $k^{mn}$.
The fundamental identity (\ref{fundi}) now reads
\be
\label{funab}
k_{mn} Q^m_i Q^n_j = 0, \;\;\;\;\; k_{mn} \widetilde{Q}^m_{\bar{\imath}} \widetilde{Q}^n_{\bar{\jmath}} = 0 .
\ee
The constraint is less restrictive than in the non-abelian case, so we have more freedom to construct new theories.
We will not try to obtain a catalogue of all possible such theories. Instead, we will consider two typical solutions to the constraint (\ref{funab}).

The first solution is to employ the linear quiver structure discussed in the non-abelian case.
The constraint (\ref{funab}) is trivially satisfied
because only one hyper-multiplet and one twisted hyper-multiplet
(and not two hyper-multiplets) meet at the same gauge group.
Unlike the non-abelian case, however, the values of the charges
can differ from $\pm 1$.

There is another simple way to satisfy the constraint (\ref{funab}).
Divide the gauge fields into two groups, say $A_m$ and $\tilde{A}_{\bar{m}}$,
and demand that $q^i$ be charged only under $A_m$ and
$\tilde{q}^{\bar{\imath}}$ only under $\tilde{A}_{\bar{m}}$.
If we further demand that $k_{mn}= 0 = k_{\bar{m}\bar{n}}$
with $k_{\bar{m}n} \neq 0$, then
the constraint (\ref{funab}) is trivially satisfied.
The Chern-Simons term in the Lagrangian now reads
\be
\CL_{\rm CS} \sim k^{\bar{m}n} \tilde{A}_{\bar{m}} d A_n .
\ee
This type of coupling is more commonly known as abelian
BF coupling ($\tilde{A} \goto B$, $dA \goto F$).
In fact, it has been known for a long time \cite{GatesBF} that abelian BF theories interacting with mater fields can be rendered $\CN=4$ supersymmetric. This is an exception (the only one
we are aware of) to the $\CN=3$ threshold mentioned
in the Introduction which predates the recent discoveries of
non-abelian $\CN \ge 4$ models.

The successes of the BL model and the GW construction
partly rely on giving up the notion of a vector multiplet.
In the absence of Yang-Mills kinetic term,
the gauge field 
does not
give rise to on-shell physical states, so it is permissible
to have only the gauge field and none of its super-partners in the Lagrangian.
In contrast, the abelian BF model of \cite{GatesBF}
maintains the whole $\CN=4$ vector-multiplet structure
and is compatible with the Yang-Mills kinetic term.

We use the following notation for the (twisted) vector multiplets:
\be
{\rm vector :}\;\;\; (A_m , \chi_m^{\da\b}, s_m^{\da\db} \; ; \; D_m^{\a\b}),
\nn \\
{\rm twisted}\; {\rm vector :}\;\;\;
(\tilde{A}_{\bar{m}} , \tilde{\chi}_{\bar{m}}^{\a\db}, \tilde{s}_{\bar{m}}^{\a\b} \; ; \; \tilde{D}_{\bar{m}}^{\da\db}),
\nn
\ee
where $\chi(\tilde{\chi})$ is the gaugino, $s(\tilde{s})$ is
the scalar triplet and $D(\tilde{D})$ is the triplet
of auxiliary fields.
The gauge index is raised and lowered by $k_{\bar{m}n}$;
for example, $\tilde{\chi}^n \equiv k^{n\bar{m}} \tilde{\chi}_{\bar{m}}$.
In the notation of the present paper, the Lagrangian of
the abelian BF theory is given by
\footnote{See section II of \cite{kapstr} for a review
of the abelian BF theories in the $\CN=2$ superspace language.}
\begin{eqnarray}
 {\cal L} &=&
 \frac{\varepsilon^{\m\n\l}}{4\pi}
 k_{\bar{m}n} \tilde{A}^{\bar{m}}_\m \partial_\n A^n_\l
 + \frac{1}{4\pi}\left( i \tilde{\chi}^n_{\a\db} \chi_n^{\db\a} + \tilde{s}^n_{\a\b} D_n^{\a\b} +\tilde{D}^n_{\da\db} s_n^{\da\db} \right)
\nn\\&&
 +\frac12\omega_{AB}
  \left(-\epsilon^{\a\b} D q^A_\a D q^B_\b
        +i\epsilon^{\da\db}\psi^A_\da \Ds\psi^B_\db \right)
 +\frac12\tilde\omega_{AB}
  \left(-\epsilon^{\da\db} D \tilde q^A_\da D \tilde q^B_\db
       +i\epsilon^{\a\b}\tilde\psi^A_\a \Ds\tilde\psi^B_\b \right)
 \nn\\&&
 -\frac{i}{2} s_n^{\da\db} (\psi\psi)^n_{\da\db}
 -\frac{i}{2} \tilde{s}_{\bar{n}}^{\a\b} (\tps\tps)^{\bar{n}}_{\a\b} -\frac14 (\mu^{mn})^\a_{\ \a} s_m^{\da \db} s_{m \da \db} -\frac14 (\tilde{\mu}^{mn})^\da_{\ \da} \tilde{s}_m^{\a \b} \tilde{s}_{m \a \b} 
 \nonumber \\ &&
 + \chi_n^{\da\b} \jmath^n_{\b\da}
 + \tilde{\chi}_n^{\a\db} \tilde{\jmath}^n_{\db\a} + \frac12 D_n^{\a\b} \m^n_{\a\b} + \frac12 \tilde{D}_{\bar{n}}^{\da\db} \tilde{\m}^{\bar{n}}_{\da\db} .
\label{LBF}
\end{eqnarray}
Upon integrating out $(\chi, \tilde{\chi}, s, \tilde{s}, D, \tilde{D})$,
we obtain a Lagrangian which precisely coincides
with the general Lagrangian (\ref{Lful2}) specialized
to the abelian BF assignments of CS couplings and charge matrices.
Note that, at first sight, the general Lagrangian
(\ref{Lful2}) look different from (\ref{LBF}) above,
because (\ref{Lful2}) includes both unmixed and mixed couplings but (\ref{LBF}) only allows mixed couplings.
However, note that the $\m^3/\tilde{\m}^3$ parts of the bosonic potential
in (\ref{Lful2}) vanish for abelian theories,
and the $(\jmath\jmath)$/$(\tilde{\jmath}\tilde{\jmath})$ Yukawa couplings vanish when the $k^{mn}$ is a BF-type.

Finally, we note that the two solutions to the constraints (\ref{funab}) 
are not mutually exclusive. 
In the next section, we will see some abelian theories which 
can be understood from both points of view up to a change of 
basis for the gauge fields. 

\subsection{Mass deformation}

So far, we have restricted our attention to superconformal theories.
In this subsection, we consider taking a mass deformation of the new theories.
We will focus on mass
parameters which preserve the $SO(4)$ R-symmetry.
In the GW construction procedure, the mass term is added to the
superpotential as $W=W_{0} + \Delta W + W_{\rm mass}$,
where $W_0$ and $\D W$ were defined in (\ref{w0}, \ref{w1}), while $W_{\text{mass}}$ is given by
\begin{eqnarray}
  W_{\text{mass}} = \frac m2 \epsilon^{\alpha \beta} \omega_{AB}
  q^A_\alpha q^B_\beta - \frac {m'}{2} \epsilon^{\da \db} \omega_{AB} \tilde{q}^A_{\da} \tilde{q}^B_{\db}.
  \label{wm}
\end{eqnarray}
Supersymmetry transformation rules should be modified accordingly,
$\delta \Phi=\delta_0 \Phi+ \delta_{\text{mass}} \Phi$, for various fields $\Phi$, where $\delta_0 \Phi$ are the undeformed transformation (\ref{Sful2}). 
The only non-trivial changes due to the mass deformation are
\begin{eqnarray}
  \delta_{\rm mass}\psi^A_\da = m q^A_\a \eta^\a_{\ \da}, \hspace{0.7cm}
  \delta_{\rm mass}\tilde{\psi}^A_\a =
  m' \tilde{q}^A_\da \eta_\a^{\ \da}.
\end{eqnarray}

A straightforward computation shows that
the only additional contributions
to the Lagrangian that can potentially
break the $SO(4)$ R-symmetry come from
\begin{eqnarray}
  \frac12 \left(F^A_\alpha\right)^2 &=& \cdots - 2m\pi k_{mn} \epsilon^{\alpha \da} \epsilon^{\beta \db} \mu^m_{\alpha\beta} \tilde{\mu}^n_{\da\db},
  \nonumber \\
  \frac12 \left(\tilde{F}^A_{\dot{m}}\right)^2 &=& \cdots + 2m'\pi k_{mn} \epsilon^{\alpha \da} \epsilon^{\beta \db} \mu^m_{\alpha\beta} \tilde{\mu}^n_{\da\db}.
\end{eqnarray}
It follows that the two terms cancel against each other
if and only if the two mass parameters for
the hyper and twisted hyper-multiplets are equal: $m = m'$.
In summary, the mass-deformed Lagrangian is
\begin{eqnarray}\label{massdeformedGW}
  \CL_{\rm mass} &=& -\omega_{AB} \left(\frac{m^2}{2}  \epsilon^{\a \b} q^A_\a q^B_\b + \frac{m^2}{2} \epsilon^{\da \db} \tq^A_\da \tq^B_\db + \frac{i}{2} m \epsilon^{\da \db} \psi^A_\da \psi^B_\b - \frac{i}{2} m \epsilon^{\a \b} \tilde{\psi}^A_\a \tilde{\psi }^B_\b \right) \nonumber \\ && \hspace{0.5cm} 
  - \frac{2\pi}{3} m k_{mn} \left((\mu^m)_{\a \b} (\mu^n)^{\b \a}  - (\tilde{\mu}^m)_{\da \db} (\tilde{\mu}^n)^{\db \da} \right),
\end{eqnarray}
with deformed SUSY variation rules
\begin{eqnarray}\label{massdeformedGWsusy}
  \delta_{\rm mass}\psi^A_\da = m q^A_\a \eta^\a_{\ \da}, \hspace{0.7cm}
  \delta_{\rm mass}\tilde{\psi}^A_\a =
  m \tilde{q}^A_\da \eta_\a^{\ \da}.
\end{eqnarray}
%


\section{Bagger-Lambert Theory and M-Crystal Model}

\subsection{Bagger-Lambert theory}

Bagger and Lambert \cite{BL2} recently constructed three-dimensional $\CN=8$ superconformal Chern-Simons theories, believed to describe multiple M2-branes. In their construction, some 3-algebra with the four-index structure constant $f^{abc}_{\ \ \ d}$ was introduced. It can be however shown that the only possible 3-algebra is in fact $SO(4)$ with $f^{abcd}=\ve^{abcd}$, once we suppose $h^{ab}=\text{tr}(t^at^b)$ is positive definite~\cite{x5,x6}. We will show that, from the $\CN=4$ perspective, the $SO(4)$ BL model can be regarded as the GW model of the same gauge group with additional twisted hyper-multiplet.

\paragraph{BL theory}

Let us start with a summary of our conventions for the BL theory. We use the mostly plus metric and mostly Hermitian eleven-dimensional Gamma matrices
\begin{eqnarray}
  \eta^{\mu\nu}=\text{diag}(-1, 1, \cdots, 1), \hspace{0.7cm}
  \{ \Gamma^\mu , \Gamma^\nu \} = 2\eta^{\mu\nu}.
\end{eqnarray}
Spinors $\Psi^a$ and supersymmetry parameter $\varepsilon$ are  eigenvectors of $\Gamma^{012}$ and $\Gamma^{3\cdots10}$ (we use $\Gamma^{012}\Gamma^{3\cdots10}=-1$) such that
\be
 \Gamma^{012}\Psi^a=-\Psi^a,~~~
 \Gamma^{012}\varepsilon=\varepsilon,~~~
 \Gamma^{3\cdots10}\Psi^a=\Psi^a,~~~
 \Gamma^{3\cdots10}\varepsilon=-\varepsilon. \label{condition}
\ee
Here $a$ denote $SO(4)$ gauge indices. The Lagrangian for the $SO(4)$ BL model reads
\begin{eqnarray}
{\cal L} &=&
 -\frac12D_\mu X_I^a D^\mu X_I^a
 +\frac i2\bar\Psi^a \Gamma^\mu D_\mu \Psi^a
 -\frac {i}{4\kappa}\varepsilon^{abcd}\;
  \bar\Psi^a X_I^bX_J^c\Gamma^{IJ}\Psi^d
 \nn\\&&
 +\frac{\kappa\epsilon^{\mu\nu\l}}{2}
\left(
      A_\mu^{ab}\partial_\nu\tilde A_\lambda^{ab}
     +\frac23 A_\mu^{ab} \tilde A_\nu^{ac} \tilde A_\lambda^{bc}
 \right)
 -\frac1{12\kappa^2}\sum_{IJK,a}
 (\varepsilon^{abcd}X^b_IX^c_JX^d_K)^2,
 \label{blaction}
\end{eqnarray}
and the supersymmetry transformation rules are
\begin{eqnarray}
 \delta X^a_I &=& i\bar\varepsilon\Gamma_I\Psi^a,\nn\\
 \delta\Psi^a &=&
 \sla DX^a_I\Gamma^I\varepsilon
 +\frac 1{6\kappa}\varepsilon^{abcd}X^b_IX^c_JX^d_K\Gamma^{IJK}\varepsilon, \nn\\
 \delta \tilde{A}_\mu^{ab} &=& -
 \frac{i}{\kappa}\varepsilon^{abcd}\;
 \bar\ve\Gamma_\mu\Gamma^IX_I^c\Psi^d.
\label{blsusy}
\end{eqnarray}
Hew we used the covariant derivatives and the tilded gauge field,
\[
 D\Psi^a\equiv d\Psi^a+\tilde A^{ab}\Psi^b,~~~~
 \tilde A^{ab}=\varepsilon^{abcd}A^{cd}.
\]
It is noteworthy here that the standard quantization rule of Chern-Simons coupling gives $2 \pi \kappa \in \ZZ$. In order to verify that this BL model nicely fits into the extended GW model with $PSU(2|2)$, we first reduce the number of supersymmetry by half.

\paragraph{R-symmetry representation}

In reducing $\CN=8$ supersymmetry to $\CN=4$,
it is useful to keep track of how the $SO(8)$
R-symmetry gets broken:
\be
SO(8) \supset SO(4)_1 \times SO(4)_2 \sim
\left( SU(2)_L \times SU(2)_A \right) \times \left( SU(2)_B \times SU(2)_R \right).
\ee
The two $SO(4)$ factors rotate $X_{3,4,5,6}$ and $X_{7,8,9,10}$ separately.
For clarity, let us rename $X^a_{I=7,8,9,10}$ as $Y_I^a$ from here on.
In terms of the $SU(2)$ factors, $X_I$ transform as
$(\bf{2,2,1,1})$ and $Y_I$ as $(\bf{1,1,2,2})$.
As for the spinors, define
\begin{eqnarray}\label{fermion}
  && \tilde{\psi} \equiv \frac{1+\Gamma^{3456}}{2} \Psi, \hspace{0.5cm}
  \psi \equiv \frac{1-\Gamma^{3456}}{2} \Psi
  \hspace{0.5cm}(\Gamma^{3456789\overline{10}}\Psi=\Psi),
\\
\label{susycondition}
  && \eta =\frac{1+\Gamma^{3456}}{2}\varepsilon, \hspace{0.6cm}
  \tilde{\eta} =\frac{1-\Gamma^{3456}}{2}\varepsilon
  \hspace{0.6cm}(\Gamma^{3456789\overline{10}}\varepsilon=-\varepsilon).
\end{eqnarray}
They transform under the four $SU(2)$ factors as
\be
\tilde{\psi} \; : \; (\bf{2,1,2,1}), \;\;
\psi \; : \; (\bf{1,2,1,2}), \;\;\;\;
\eta \; : \; (\bf{2,1,1,2}), \;\;
\tilde{\eta} \; : \; (\bf{1,2,2,1}).
\ee

Truncation from $\CN=8$ to $\CN=4$ amounts to setting $\tilde{\eta}=0$. It is clear that, among the four $SU(2)$ factors, $SU(2)_L \times SU(2)_R$
becomes the $SO(4)$ R-symmetry of $\CN=4$.
It is also clear from the supersymmetry variation rule ($X \sim \bar{\ve} \G \Psi$)
that $X_I$ and $\psi$ form a hyper-multiplet $q$, and $Y_I$ and $\tilde{\psi}$ form a twisted hyper-multiplet $\tq$. 

As a short comment, one can truncate the $\CN=8$ BL model consistently down to $\CN=4$ by dropping the twisted hyper-multiplet: the truncated Lagrangian becomes
\begin{eqnarray}\label{BL4}
  {\cal L} &=& -\frac12 D_\mu X_{I}^a D^\mu X_{I}^a + \frac i2
  \psi^a \Gamma^\mu D_\mu \psi^a - \frac{i}{4\kappa}\ve^{abcd}
  \bar{\psi}^a \Gamma_{IJ} \psi^b X_{I}^c X_{J}^d \nonumber \\&&
 +\frac{\kappa\epsilon^{\mu\nu\l}}{2}
\left(
      A_\mu^{ab}\partial_\nu\tilde A_\lambda^{ab}
     +\frac23 A_\mu^{ab} \tilde A_\nu^{ac} \tilde A_\lambda^{bc}
 \right)
 -\frac1{12\kappa^2}\sum_{IJK,a}
 (\varepsilon^{abcd}X^b_IX^c_JX^d_K)^2,
\end{eqnarray}
where $I$ here runs over 3 to 6. The truncated 
supersymmetry transformation rules are
\begin{eqnarray}
  \delta X_I^a &=& i\bar{\eta} \Gamma_I \psi^a \nonumber \\
  \delta \psi^a &=& \left( D_\mu X_I^a \Gamma^\mu \Gamma^I  +
  \frac{1}{6\kappa} \ve_{abcd}  \Gamma^{IJK} X_I^b X_J^c X_K^d
  \right) \eta \nonumber \\
  \delta \tilde{A}_\mu^{ab} &=& -
 \frac{i}{\kappa}\varepsilon^{abcd}\;
 \bar\eta\Gamma_\mu\Gamma^IX_I^c\psi^d.
\end{eqnarray}
%

%
%

\paragraph{Explicit embedding}

{}For explicit comparison, we will use the doublet indices
\[
(\a,\b; \dot{\s}, \dot{\t}; \s, \t; \da, \db)
\]
for the four $SU(2)$ R-symmetry factors. For each $SU(2)$, indices are raised and lowered by the invariant anti-symmetric tensor satisfying $\ve_{\a\g} \ve^{\g\b} = \d_\a{}^\b$.
The pseudo-reality condition for a ``spinor'' reads
\[
(u_\a)^\dagger = \bar{u}^\a = \e^{\a\b} u_\b .
\]
Explicit embedding of the $SU(2)$ factors into the $SO(8)$ 
is facilitated by
a specific basis for the eleven-dimensional gamma matrices,
\begin{eqnarray}
  \Gamma^\mu &=& \gamma^\mu \otimes (-\gamma^5) \otimes \gamma^5 \hspace{0.4cm} \text{ for } \mu=0,1,2   \nonumber \\
  \Gamma^I  &=& {\bf 1}_2 \otimes
  \gamma^I \otimes {\bf 1}_4 \hspace{1cm} \text{ for
  }I=3,4,5,6 \nonumber \\
  \Gamma^J &=& {\bf 1}_2 \otimes \gamma^5 \otimes \tilde{\gamma}^J \hspace{0.95cm}
  \text{ for }J=7,8,9,10
\end{eqnarray}
with hermitian $SO(4)$ gamma matrices
\begin{eqnarray}
  \gamma^I = \begin{pmatrix} 0 & (e^I)_{\a\dot{\t}} \\ (\bar{e}^I)^{\dot{\s}\b} & 0
  \end{pmatrix}, 
\end{eqnarray}
and the real three-dimensional gamma matrices $\gamma^\mu$. These $\gamma^\mu$ are chosen to satisfy $\gamma^{012}=1$. Here $e^I=(i\vec{\s},1)$ and $\bar{e}^I=(-i\vec{\s},1)$ satisfy the reality condition
\be
\label{reale}
(e^I_{\a\dot{\s}})^* = \e^{\a\b} (e^I)_{\b\dot{\t}} \e^{\dot{\t}\dot{\s}} .
\ee
The other four gamma matrices $\tilde{\g}^J$ are numerically identical to $\g^I$, but of course carry different $SU(2)$ indices. In this basis, the reality condition (Majorana condition) on fermion fields $\Psi$ is given by
\begin{eqnarray}\label{Mcondiion}
  \Psi^* = B \Psi, \hspace{1cm} B=\Gamma^{3579}={\bf 1}_2 \otimes C \otimes C,
\end{eqnarray}
where $C$ denote the charge-conjugation operator
\begin{eqnarray}
  C=\begin{pmatrix} \epsilon & 0 \\ 0 & \epsilon^{-1} \end{pmatrix}.
\end{eqnarray}
Decomposing the spinors as
\be
\Psi ~ \goto ~  \tps_{\a\s} \oplus \psi^{\dot{\s}\da}, \;\;\;\;\;
\ve ~ \goto ~ \eta_\a^{\ \da} \oplus \tilde{\eta}^{\dot{\s}}_{\ \s},
\ee
one can show that the reality conditions (\ref{Mcondiion}) become
\be
(\tps_{\a\s})^* = \e^{\a\b} \e^{\s\t} \tps_{\b\t}, \;\; {\rm etc.}
\ee
For later convenience, let us re-express the scalars  as bi-spinors via
\be
X_I \;\; \goto \;\; \bar{X}^{\dot{\s}\a} \equiv \frac12 X_I (\bar{e}^I)^{\dot{\s}\a}, \;\; {\rm etc.} \; ,
\ee
whose reality condition can be read off from (\ref{reale}). 
After all the replacements, the Lagrangian can now be described as

\begin{eqnarray}\label{n4BL2}
  \CL &=&  \text{tr} ( - D_\mu \bar{X}^a D^\mu X^a - D_\mu \bar{Y}^a D^\mu Y^a + \frac{i}{2} \psi^a \gamma^\mu D_\mu \psi^a
  + \frac{i}{2} \tilde{\psi}^a \gamma^\mu D_\mu \tilde{\psi}^a) \nonumber \\  &&\hspace{-0.7cm} -\frac{i}{\kappa} \ve^{abcd} \left( \psi^a_{\ \dot{\s} \da} (\bar{X}^c X^d)^{\dot{\s}}_{\ \dot{\t}} \psi^{b\dot{\t}\da} + \tilde{\psi}^{a\a\s} (Y^c \bar{Y}^d)_\s^{\ \t}  \tilde{\psi}^b_{\ \a\t} - 4 \psi^a_{\ \dot{\s} \da} \bar{X}^{c \dot{\s} \a} \bar{Y}^{d \da \s} \tilde{\psi}^b_{\a \s} \right)  \nonumber \\ && \hspace{-0.7cm}
 +\frac{\kappa\epsilon^{\mu\nu\l}}{2}
\left(
      A_\mu^{ab}\partial_\nu\tilde A_\lambda^{ab}
     +\frac23 A_\mu^{ab} \tilde A_\nu^{ac} \tilde A_\lambda^{bc}
 \right)
 -V(X,Y),
\end{eqnarray}
with scalar potential $V=V_1(X) + V_1(Y) + V_2(X,Y) + V_2(Y,X)$ with
\begin{eqnarray}
V_1(X)&=& -\frac{4 }{9\kappa^2}\ve_{abcd} \ve_{aefg} \text{tr}\left( X^b \bar{X}^c X^d \bar{X}^e X^f \bar{X}^g \right), \nonumber \\
V_2(X,Y) &=& -\frac{2}{\kappa^2} \ve_{abcd} \ve_{aefg} \text{tr} \left( X^b \bar{X}^e \right) \text{tr} \left( Y^c \bar{Y}^d Y^f \bar{Y}^g \right).
\end{eqnarray}
The supersymmetry transformation rules (\ref{blsusy}) can be recast as
\begin{eqnarray}\label{susyn8bl}
  \delta X^a_{\alpha\dot{\s}} &=&
  i \eta_\alpha^{\ \da} \psi^{a}_{\ \dot{\s} \da},
  \hspace{0.7cm} \delta Y^a_{\ \s\da} ~=~ -i \tilde{\psi}^a_{\  \a \s} \eta^\a_{\ \da},
  \nonumber \\
  \delta \psi^a_{\ \dot{\s} \da} &=& \left[ - 2 \gamma^\mu D_\mu X^a_{\ \a \dot{\s} } +  \frac{8\pi}{3n} \ve_{abcd}  ( X^b \bar{X}^c X^d )_{\alpha \dot{\s} } \right] \eta^\alpha_{\ \da}
  + \frac{8\pi}{n} \ve_{abcd} X^b_{\ \a \dot{\s}} \eta^\a_{\ \db} (\bar{Y}^c Y^d)^{\db}_{\ \da} , \nonumber \\
  \delta \tilde{\psi}^a_{\ \alpha \s} &=& \left[
  2 \gamma^\mu D_\mu Y^a_{\s\dot{\a}} + \frac{8\pi}{3n} \ve^{abcd}
  ( Y^b \bar{Y}^c Y^d )_{\s \dot{\a}} \right] \eta_\alpha^{\ \dot{\a}}
  + \frac{8\pi}{n} \ve^{abcd} Y^b_{\s\dot{\a}}
  (X^c \bar{X}^d)_\alpha^{\ \beta} \eta_\beta^{\ \dot{\a}} ,
  \nonumber \\
  \delta \tilde{A}_\mu^{ab} &=&  - \frac{4 \pi i}{n} \ve_{abcd} \eta^\a_{\ \da} \gamma_\mu \left(   \psi^{d \dot{\s} \da}X^c_{\ \a \dot{\s}} + \tilde{\psi}^d_{\ \a \s} \bar{Y}^{c \da \s}  \right).
\end{eqnarray}
where we used the fact that $\kappa$ is quantized as $2\pi
\kappa=n$ ($n$ is integer). Before the detailed comparison, let us first present the convenient choice of symplectic embedding of $SO(4)_{\rm gauge}$.

\paragraph{Symplectic Embedding}

For the $PSU(2|2)$ GW theory, the matters transform as bi-fundamentals under the gauge group $SU(2)\times SU(2)$, or as vector under $SO(4)_{\text{gauge}}$. For clarity, we will use the latter convention throughout this section. The representation under
$SO(4)_{\rm gauge} \times SU(2)_A \times SU(2)_B$ symmetry, aside from the $SU(2)_L\times SU(2)_R$ R-symmetry, of the BL model can be related to the symplectic notation of Gaiotto and Witten as follows: a symplectic index $A$  on hyper-multiplets can be decomposed into $(a,\dot{\s})$ while $B$ on twisted ones can be decomposed into $(b,\s)$.  Here $a,b$ denote the $SO(4)_{\text{gauge}}$ index, and $\dot{\s}$ and $\s$ transform as doublets under $SU(2)_A$ and $SU(2)_B$, respectively.

Under this symplectic embedding of $SO(4)_{\text{gauge}}$ into $Sp(8)$, the symplectic indices simplifies the index structure of various matter fields as
\begin{eqnarray}
  q^A_\alpha &=& \sqrt2 X^a_{\ \a \dot{\s}}, \hspace{0.5cm}
  \tilde{q}^A_{\da} ~=~ \sqrt2 Y^a_{\ \s \da}, \nonumber \\
  \psi^A_\da &=& \psi^a_{\ \dot{\s} \da}  \hspace{0.5cm} \tilde{\psi}^A_\a ~=~  \tilde{\psi}^a_{\ \a\s}
\end{eqnarray}
The symplectic invariant tensor and the gauge generators are then decomposed as
\begin{eqnarray}
  \omega_{AB} = \delta_{ab}\otimes \epsilon^{\dot{\s}\dot{\t}}\ ,
\;\;\;\;\;
  t^m_{AB} = t^m_{ab} \otimes \epsilon^{\dot{\s}\dot{\t}}  ,   \label{sym1}
\end{eqnarray}
with $SO(4)_{\rm gauge}$ group generators $t^m_{ab}$. Clearly, the same argument goes through for the twisted hyper-multiplets:
\begin{eqnarray}
  \tilde{\omega}_{AB} = \delta_{ab}\otimes \epsilon^{\s\t}.
\;\;\;\;\;
  \tilde{t}^m_{AB} = t^m_{ab}\otimes \epsilon^{\s\t}. \label{sym2}
\end{eqnarray}
We are now ready to show the equivalence between the BL model and the extend GW model with the supergroup $PSU(2|2)$.

\paragraph{Equivalence}

Re-normalizing the supersymmetry parameters $\eta$ as
$
 \eta_\alpha^{\ \da}  \rightarrow \frac{1}{\sqrt2} \eta_\a^{ \ \da},
$
together with the choice of symplectic embedding above, one can obtain from (\ref{susyn8bl}) the following supersymmetry variation rules
\begin{eqnarray}
  \delta q^A_\a &=& i \eta_\a^{\ \da} \psi^A_\da,
  \hspace{0.7cm}  \tq^A_\da ~=~ -i\tilde{\psi}^A_\a \eta^\a_{\ \da},  \nonumber \\
  \delta \psi^A_{\  \da} &=&  \left[ \gamma^\mu D_\mu q^A_\a + \frac{2\pi}{3} k_{mn} (t^m)^A_{\ B} q^B_\b (\mu^n)^\b_{\ \a} \right] \eta^\a_{\ \da} - 2\pi k_{mn} (t^m)^A_{\ B} q^B_\a (\tilde{\mu}^n)_\da^{\ \db} \eta^\a_{\ \db}  \nonumber \\
  \delta \tilde{\psi}^A_{\ \alpha} &=& - \left[ \gamma^\mu D_\mu q^A_\a + \frac{2\pi}{3} k_{mn} (\tilde{t}^m)^A_{\ B} \tq^B_\db (\tilde{\mu}^n)^\db_{\ \da} \right] \eta_\a^{\ \da} + 2 \pi k_{mn} (\tilde{t}^m)^A_{\ B} \tilde{q}^B_\da (\mu^n)^\b_{\ \a} \eta_\b^{\ \da} \nonumber \\ \delta \tilde{A}^m_\mu &=&  2\pi i \eta^{\a\da} \rho_\mu  (q^A_\a \tau^m_{AB} \psi^{B}_\da  - \tilde{q}^A_\da \tilde{\tau}^m_{AB} \tilde{\psi}^B_\a) ,
  \label{Sbl3}
\end{eqnarray}
where $\tilde A_\mu$ is defined by
$
  \tilde{A}_\mu^{ab} = k_{mn} (t^m)_{ab} \tilde{A}^n_\mu 
$
and we chose a basis for $SO(4)_{\text{gauge}}$ such that
\begin{eqnarray}
  (t^m)_{ab} &=& (\vec{t}_L, \vec{t}_R)_{ab}, \hspace{1cm} k^{mn}
  = k \; \text{diag}({\bf 1}_3, -{\bf 1}_3),
\end{eqnarray}
where $\vec{t}_L$ and $\vec{t}_R$ denote self-dual and anti-self-dual $SO(4)_{\text{gauge}}$ generators. Here $k$ now is quantized to be an even integer.  A useful identity in our discussion is
\begin{eqnarray}
  k_{mn} (t^m)_{ab} (t^n)_{cd} = \frac{2}{k} \varepsilon_{abcd}.
\end{eqnarray}

The transformation rules (\ref{Sbl3}) are in perfect agreement with those in (\ref{Sful2}) of the extended GW model in section 2, which strongly implies the equivalence between the two models. One can also easily show that the Lagrangian (\ref{n4BL2}) reduces to (\ref{Lful2}). 

In summary, we conclude that the BL model in the $\CN=4$ notation is identical to the extended GW model with the supergroup $PSU(2|2)$.

\paragraph{Mass deformation}

We now show that the mass-deformed BL model proposed in \cite{x3,HLL08} is in fact
a special example of the mass-deformed extended GW model in section 2.3.
The Lagrangian of the mass-deformed BL model can be written as
$\CL = \CL_{\text{BL}} + \CL_{\text{mass}}$ with
\begin{eqnarray}\label{massdeformedBL}
  \CL_{\text{mass}} = -\frac{m^2}{2} (X_I^a)^2 - \frac{i}{2}m \bar{\Psi}^a \Gamma_{3456} \Psi^a + \frac{4m}{\kappa} \ve^{abcd} (X_3^a X_4^b X_5^c X_6^d + X_7^a X_8^b X_9^c X_{10}^d),
\end{eqnarray}
which is invariant under the deformed supersymmetry transformation rules, 
$\delta \Phi = \delta_{\rm BL} \Phi + \delta_{\rm mass} \Phi$, for various fields $\Phi$ with
\begin{eqnarray}
  \delta_{\rm mass} \Psi^a = m \Gamma_{3456} X_I^a \Gamma^I \epsilon.
\end{eqnarray}
This deformation preserves the maximal, say sixteen real, supersymmetries. Let us rewrite them in the $\CN=4$ language. Based on the specific representation of eleven-dimensional gamma matrices together with the choice of symplectic embedding in this section, the deformed supersymmetry 
variation rules now becomes
\begin{eqnarray}\label{masssusy}
  \delta_{\rm mass} \psi^a_{\ \dot{\s} \da} = 2m X^a_{\ \a \da} \eta^\a_{\ \da} \ &\rightarrow& \ \delta_{\rm mass} \psi^A_\da = m q^A_\a \eta^a_{\ \da} \nonumber \\
  \delta_{\rm mass} \tilde{\psi}^a_{\ \a \s} = 2m Y^a_{\ \s \da} \eta_\a^{\ \da} \ &\rightarrow& \ \delta_{\rm mass} \tilde{\psi}^A_\a = m \tilde{q}^A_{\da}  \eta_\a^{\ \da},
\end{eqnarray}
which agrees with (\ref{massdeformedGWsusy}). It implies the equivalence between the two models.
One can also show that the interaction terms in (\ref{massdeformedBL}) reduce to those in (\ref{massdeformedGW}).


\subsection{Abelian quiver and M-crystal model}

\paragraph{Review of the M-crystal}

\begin{figure}[htbp]
\begin{center}
\includegraphics[width=11cm]{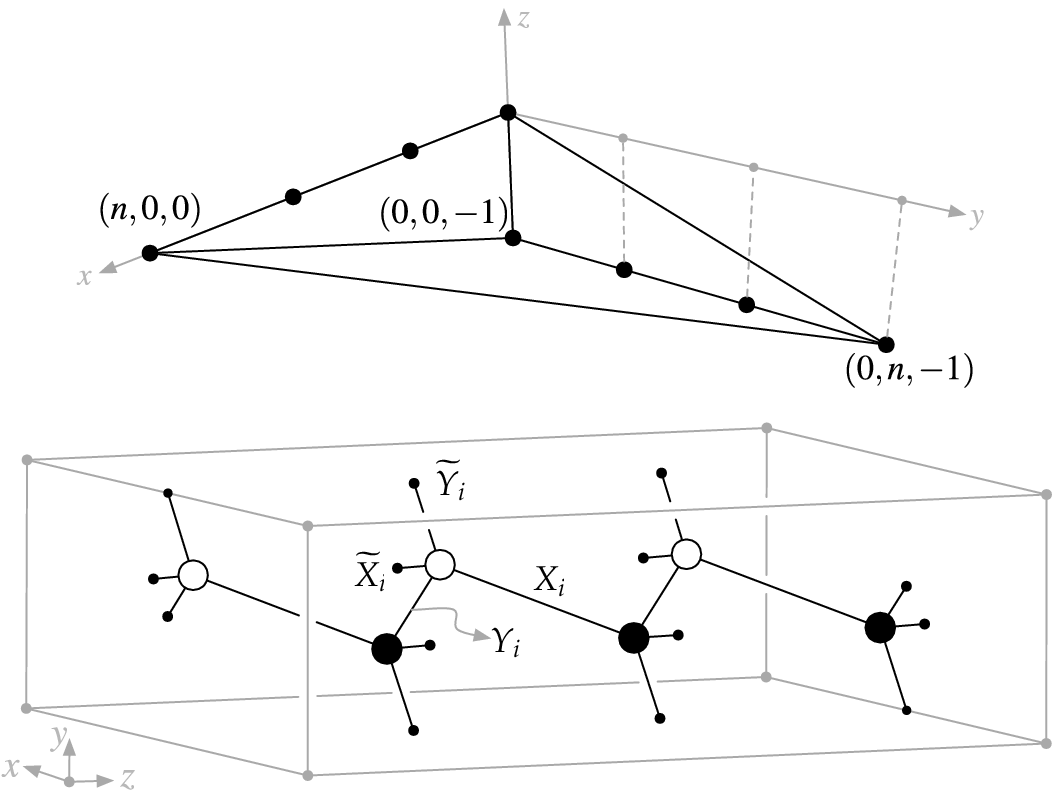}
\caption{Toric diagram and crystal for $(\IC^2/\IZ_n)^2$ (reproduced from
\cite{mc3}).} \label{znzn}
\end{center}
\end{figure}

The M-crystal model for M2-branes probing the $(\IC^2/\IZ_n)^2$ orbifold was studied in \cite{mc3}. The toric diagram of the orbifold and the associated crystal diagram are reproduced in Figure \ref{znzn}.

In the crystal diagram, the ``bonds'' correspond
to the $\CN=2$ holomorphic matter super-fields.
The ``atoms'' of the crystal encode the super-potential;
take the products of all fields ending on a given atom
and sum over all atoms with opposite signs between
the white and black atoms. The result is
\begin{eqnarray}
\label{wzn}
  W = \sum_{i=1}^{n} \left( X_i \widetilde{X}_i
  Y_i \widetilde{Y}_i -  X_i \widetilde{X}_i
  Y_{i+1} \widetilde{Y}_{i+1} \right),
\end{eqnarray}
from which we find the F-term conditions,
\begin{eqnarray}
 X_i \widetilde{X}_i - X_{i-1} \widetilde{X}_{i-1} = 0, \;\;\;\;\;
 Y_i \widetilde{Y}_i - Y_{i+1} \widetilde{Y}_{i+1}  = 0.
 \label{fterm}
\end{eqnarray}

In a more delicate but systematic way explained in \cite{mc3},
the crystal diagram also determines the charges of
the abelian probe theory.
In the case at hand, the gauge group turns out to be
$\left(U(1)^n/U(1)_D\right)_X
\times \left(U(1)^n/U(1)_D\right)_Y$.
The matter fields $(X_i , \widetilde{X}_i)$ are charged only under the first factor,
$(U(1)^n/U(1)_D)_X$, as
\begin{center}
\begin{tabular}{l|cc}
  & $X_i$ & $\widetilde{X}_i$   \cr \hline
$Q_i$ & + & $-$  \cr
$Q_{i+1}$ & $-$ & +
\end{tabular}
\hskip 1cm
($i=1, \cdots, n$)
\end{center}
with all other charges vanishing.
The matter fields
$(Y_i , \widetilde{Y}_i)$ are charged under the second factor, $(U(1)^n/U(1)_D)_Y$, in the same way.

The proposal for the abelian theory in \cite{mc3} was incomplete
as the gauge kinetic terms were not specified.
It was assumed that the D-term potential exists
and effectively complexifies the gauge group, as is 
commonly true of gauge theories with four supercharges.
Under these assumptions, the moduli space of vacua was shown to
coincide with the orbifold $(\IC^2/\IZ_n)^2$.
In fact, the orbifold theories have $\CN=4$ supersymmetry
with the $\CN=2$ fields paring up to form (twisted) hyper-multiplets,
\be
(X_i , \widetilde{X}_i^\dagger) \goto q_i, \;\;\;
(Y_i , \widetilde{Y}_i^\dagger) \goto \tilde{q}_i,
\ee
so the F-term and D-term conditions together are expected to result in a hyper-K\"ahler quotient \cite{K} for the moduli space of vacua. 

We illustrate the identification of the moduli space of vacua 
using the simplest $(\IC^2/\IZ_2)^2$ orbifold. The theory has two hyper-multiplets $(X_i,\tilde{X}_i)$ ($i=1,2$) and two twisted hyper-multiplets $(Y_i,\tilde{Y}_i)$. Their charger are summarized in the following table, 
\begin{table}[htb]
\begin{center}
\begin{tabular}{r|cccccccc}
  & $X_1$ & $\tilde{X}_1$ & $X_2$ & $\tilde{X}_2$ & $Y_1$ & $\tilde{Y}_1$ &  $Y_2$ & $\tilde{Y}_2$ \cr
  \hline
$Q^X$ & $-$ & $+$ & $-$ & $+$ & $0$ & $0$ & $0$ & $0$ \cr
$\tilde{Q}^X$ & $+$ & $-$ & $+$ & $-$ & $0$ & $0$ & $0$ & $0$ \cr
$Q^Y$ & $0$ & $0$ & $0$ &$ 0$ & $+$ & $-$ & $+$ & $-$ \cr
$\tilde{Q}^Y$ & $0$ & $0$ & $0$ &$ 0$ & $-$ & $+$ & $-$ & $+$
\end{tabular}
\end{center}
\end{table}\vspace{-0.5cm}

\noindent
and the superpotential is given by 
\begin{eqnarray}
  W= (X_1 \tilde{X}_1 - X_2 \tilde{X}_2)(Y_1 \tilde{Y}_1 - Y_2 \tilde{Y}_2).
\end{eqnarray}
The F-term conditions require that $X_1 \tilde{X}_1 = X_2 \tilde{X}_2 \equiv u_3$ and $Y_1 \tilde{Y}_1 = Y_2 \tilde{Y}_2 \equiv v_3$. The gauge invariant monomials $u_1=X_1 \tilde{X}_2$, $u_2=\tilde{X}_1 X_2$, $v_1 = Y_1 \tilde{Y}_2$ and $v_2=\tilde{Y}_1 Y_2$ together with $u_3$ and $v_3$ parameterize the moduli space of vacua. Based on the F-term conditions, 
we find 
\begin{eqnarray}
  u_1 u_2 = u_3^2, \hspace{0.7cm} v_1 v_2 = v_3^2,
\end{eqnarray}
which are nothing but the algebraic descriptions 
for each of the two $(\IC^2/\IZ_2)$ factors.

In Ref.~\cite{mc3}, it was not known how the abelian theory 
can be completed in a manifestly $\CN=4$ manner.
We will show that a particular abelian quiver theory of
the present paper satisfies all
the properties of such a completion.


\paragraph{The connection}

Consider the abelian $\hat A_{2n-1}$  quiver theory, which has $2n$ nodes connected to make a circle. We assign
\begin{itemize}
\item
A $U(1)$ gauge field $A_m$ ($m=1, \cdots, 2n$) to each node.
\item
A hyper-multiplet $q_{i}$  to each of the $n$ links
$\vev{2i-1,2i}$ $(i=1,\cdots,n)$.
\item
A twisted hyper-multiplet $\tilde{q}_{i}$ to each of the $n$ links
$\vev{2i,2i+1}$ $(i=1,\cdots,n)$.
\end{itemize}
The hyper and twisted hyper-multiplets give two ways to uplift
the group $U(1)^{2n}$ to $U(1|1)^{n}$.
The quiver diagram encodes the charges of the matter fields.
The GW construction requires the Chern-Simons coupling to be
$$
k^{mn}= k \, \text{diag}\underbrace{(1,-1,\cdots,1,-1)}_{2n}.
$$
For this model, the scalar potential as specified in (\ref{Lful2})
can be simplified as
\begin{eqnarray}\label{scalarpotential}
  V = 2\pi^2 \sum_{i=1}^{n} \left[ |\tilde{q}_{i}|^2 (\mu^{2i-1} - \mu^{2i+1})^2 + |q_{i}|^2 (\tilde{\mu}^{2i-2} - \tilde{\mu}^{2i})^2   \right],
\end{eqnarray}
where the squares of the moment maps are defined by
$(\m)^2 \equiv \m_{\a\b} \m^{\a\b}$ and $\tilde{\m}^2 \equiv \tilde{\m}_{\da\db} \tilde{\m}^{\da\db}$.
The vacuum conditions from the scalar potential become
\begin{eqnarray}\label{vacuum2}
  \mu^1 = \mu^3 =\cdots =\mu^{2n-1}, \hspace{0.5cm} \tilde{\mu}^2 = \tilde{\mu}^4 =\cdots =\tilde{\mu}^{2n} ,
  \label{vaccon}
\end{eqnarray}
which coincide with the $\CN=4$ covariant version of the F-term conditions
(\ref{fterm}).

To compare the gauge symmetries here with those
of the M-crystal model, we make the following change of basis
for the charges,
\begin{eqnarray}
  && \hspace{1cm} Q^+ =\sum_{m=1}^{2n} Q^m, \ \ \
  Q^- = \frac12 \sum_{i=1}^n (Q^{2i-1} - Q^{2i}),
  \nn \\
  &&  \hat{Q}^i = Q^{2i-2} + Q^{2i-1} - \frac{1}{n} Q^+, \ \
    \check{Q}^i = Q^{2i-1} + Q^{2i}   - \frac{1}{n} Q^+,
\end{eqnarray}
and define a new set of $U(1)$ gauge fields
\begin{equation}
 \sum_{m=1}^{2n}A_mQ^m ~=~
 A_+Q^++A_-Q^- + \sum_j\hat A_j\hat Q^j + \sum_j\check A_j\check Q^j.
\end{equation}
Recalling our charge assignments
\[
 Q^{2i-1}[q_i]=1,~~~~Q^{2i}[q_i]=-1,~~~~
 Q^{2i}[\tilde q_i]=1,~~~~Q^{2i+1}[\tilde q_i]=-1,
\]
we immediately see that $q_i$ are charged under $\hat{Q}^j$
in the same way as in the crystal model and neutral under $\check{Q}^j$.
Similarly, $\tilde q_i$ are charged under $\check{Q}^j$ and neutral
under $\hat{Q}^j$.
Note that the number of independent $\hat{Q}^j\;(\check{Q}^j)$
are $(n-1)$ as in the crystal model.
One can also show from the vacuum conditions (\ref{vacuum2}) that the moment maps associated to $\hat{A}_i,\check{A}_i$ gauge groups should vanish
\begin{eqnarray}\label{vacuum}
  \check{\mu}^i_{\a \b} = 0 \hspace{0.7cm}\hat{\mu}^i_{\da \db} = 0 \hspace{0.3cm}.
\end{eqnarray}

All matter fields are neutral under $A_+$, so $A_+$ decouple
from the dynamics of matter fields. Under $Q^-$,
all $q_i$ have charge $+1$, and all $\tilde{q}_i$ have $-1$.
As for the Chern-Simons coefficient, there are BF type couplings
between $\hat{A}_i$ and $\check{A}_j$, another BF
coupling between $A_-$ and $A_+$, and all other couplings vanish:
\begin{eqnarray}
  \CL_{\rm CS} = \frac{2n k}{4\pi }
  \epsilon^{\mu\nu\rho}A_{-\mu} \wedge \partial_\nu A_{+\rho} + \cdots,
\end{eqnarray}
where we are only keeping the BF-coupling of our interest in the following discussion.\footnote{Ref.~\cite{y3} discusses orbifolding 
the $SO(4)$ BL theory to obtain an $U(1)^2$ gauge theory, 
which resembles our $(\IC^2/\IZ_2)^2$ theory.}

Aside from the fact that the $U(1)_{Q^-}$ symmetry is gauged,
the vacuum conditions (\ref{vacuum}) together with gauge symmetries
of ($\hat{A}_i$, $\check{A}_i$) can be understood as the standard
hyper-K\"{a}hler quotient construction for the orbifold $(\IC^2/\IZ_n)^2$.
If we further mod it out by the gauge symmetry of $Q^-$,
it appears that we are left with a seven-dimensional moduli space,
in contradiction to the well-established link
between $\CN=4$ supersymmetry and hyper-K\"{a}hler spaces.

A similar problem was encountered in \cite{Lambert:2008et,Distler:2008mk},
where there was also a problematic $U(1)$ gauge field (call it $A_-$)
and a decoupled $U(1)$ gauge field (call it $A_+$)
coupled to each other through a BF coupling.
The naive result was a strange 15-dimensional moduli space of vacua.
The authors of \cite{Lambert:2008et} resolved the problem
by dualizing $A_+$ to a scalar $\sigma$ and fixing the
problematic  $A_-$ gauge symmetry by setting $\partial\sigma\sim A_-$.
The idea can be carried over to our current setup.

Starting with the Lagrangian for the present abelian quiver theories
\begin{eqnarray}
  \CL_{\rm quiver} =
 -\half(|D_\mu q_i|^2  + |D_\mu \tilde{q}_i|^2 )
 + \frac{nk}{2\pi }\epsilon^{\mu\nu\rho} A_{-\mu}\wedge\partial_\nu A_{+\rho}
 + \cdots - V(q_i,\tilde{q}_i) ,
\end{eqnarray}
we replace the decoupled $U(1)$ gauge field $A_+$ by its dual scalar $\s$ through introducing the well-known Lagrange multiplier
\begin{eqnarray}
  \CL_{\rm dual} = \frac{1}{4\pi} \epsilon^{\mu\nu\rho}
  \s \partial_\mu F_{+\nu\rho}.
\end{eqnarray}
(If we allow nonabelian embedding, magnetic monopoles would be present and the variable $\sigma$ would be
periodic.)
Here the covariant derivatives of matters are
\begin{eqnarray}
  D_\mu q_i = (\partial_\mu -iA_{-\mu} + \cdots ) q_i, \hspace{0.5cm} D_\mu \tilde{q}_i = (\partial_\mu +iA_{-\mu} + \cdots ) \tilde{q}_i, \ \ \text{ etc.},
\end{eqnarray}
where we suppressed all terms irrelevant for our discussions. It is noteworthy again that the decoupled $A^+_\mu$ enters into the theories 
through the BF-coupling with $A^-_\mu$. Collecting all interactions, one finally obtain
\begin{eqnarray}
  \CL_{\rm tot} =
  -\half(|D_\mu q_i|^2  + |D_\mu \tilde{q}_i|^2 )
  +\frac{1}{4\pi}\epsilon^{\mu\nu\rho} ( nk A_{-\mu} - \partial_\mu\s)
  F_{+\nu\rho} + \cdots  \  .
\end{eqnarray}
Note that   ${\cal L}_{\rm tot}$  is invariant under the gauge transformation  generated by  $Q^-$, which is
\begin{eqnarray}
  q_i \rightarrow e^{i\th} q_i, \hspace{0.3cm}
  \s \rightarrow \s + nk\th, \hspace{0.3cm}
  A_- \rightarrow A_- + d \th .
\end{eqnarray}
One can fix the problematic gauge field $A_{-\mu}$
by eliminating the Lagrange multiplier $F_{+\nu\rho}$
\begin{eqnarray}
  A_{-\mu} = \frac{1}{nk} \partial_\mu \s.
\end{eqnarray}
In terms of the local gauge invariant fields $ e^{i\sigma/nk} q_i$, one could   have
a possibility of  further discrete identification depending on $k$ if $\sigma$ is periodic.

To conclude, with the field redefinition $q_i \rightarrow e^{-\frac{i}{nk}\s} q_i$, the theory effectively decouples from $A_-$ and 
recovers all the desired properties of the abelian M-crystal model 
discussed earlier, except the possible discrete identification.  



\section{Discussion}

We have constructed a large class of new three-dimensional  $\CN=4$ superconformal Chern-Simons field
theories with both types of hyper-multiplets and their mass deformations, which are generically a
linear quiver theory. We find that the BL theory and the M-crystal models fit well with our theories.
  It remains to study physical properties of these theories
such as their moduli space of vacua and BPS states.
We leave them for future works. There are a few other directions which   deserve further study.

The Gaiotto-Witten construction and its extension of the present paper probably
exhaust all possible ${\cal N}=4$ superconformal Chern-Simons theories minimally coupled
to (twisted) hyper-multiplets. In Ref.~\cite{gw}, the construction 
was generalized to a sigma model whose target space is any 
hyper-K\"ahler manifold. We expect our extended 
construction will also allow such generalization.

A new superconformal family of BL models as a possible theory on $M2$ branes
were constructed quite recently where
the positive definiteness of the group space is relaxed and so the 3-algebra  structure can
be associated with arbitrary ordinary Lie algebras~\cite{bf1, bf2, bf3}. This theory has
unconventional BF couplings and the Gaiotto-Witten's work and our extension can be
 explored along this direction too. It remains to be seen how useful this further extension, which
does not fall into the classification of this paper, would be.

In the original GW construction \cite{gw},
the appearance of $U(N|M)$ or $OSp(N|M)$ theories
were related to  of
D3(O3)-branes intersecting with NS5-branes in IIB string theory.
It is natural to ask whether similar interpretation is possible for our present work
which  includes twisted hyper-multiplets.  To see this, let us explore  the following alignments of branes:
\begin{equation}
\begin{array}{l|ccccccccccc}
  & \; 0 \; & \; 1\;  & \; 2\;  & \; 3\; & \; 4\; & \;5\; & \;6\; & \;7\; & \;8\; & \;9\;     \cr \hline
{\rm D3(O3)}\; & \circ & \circ & \circ & \circ & & & & & &  \cr
{\rm NS5} & \circ & \circ & \circ & & \circ & \circ & \circ & & & \cr
{\rm D5}  &  \circ & \circ & \circ & & & & & \circ & \circ & \circ
\end{array}
  \nonumber
\end{equation}
Each of D3- and NS5-branes   breaks half of   supersymmetries,
leading to   eight real supersymmetries.  Additional
  D5-branes introduced above do not break supersymmetry
any further. In the above brane setting, the  $SU(2)_L \times SU(2)_R$
 R-symmetry is realized as
$SO(3)_{456} \times SO(3)_{789}$.
Since the hyper-multiplet  in the original Gaiotto-Witten
construction is associated to   the presence of NS5-branes,
 D5 branes would lead to the appearance of    twisted hyper-multiplets.
If such a suggestion works out correctly,
the quiver diagram of Figure \ref{quiver} would correspond to
the following brane configuration.

\begin{figure}[htbp]
\begin{center}
\includegraphics[width=8cm]{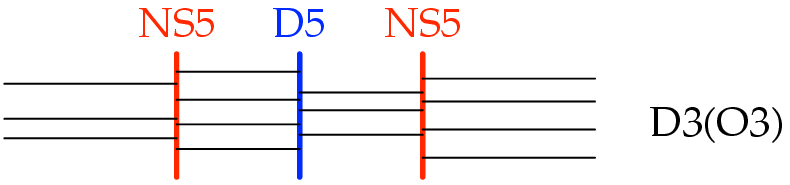}
\caption{Brane configuration
for the extended GW construction.} \label{GW-brane}
\end{center}
\end{figure}

The above argument suggests that our quiver theory have a type IIB origin.
As our theory  seems to describe the physics on M2 branes on the tip of the
 $(\IC^2/\IZ_n)^2$ orbifold, we want to find some duality transformation
 between the two approaches.  Now, replace each  $\IC^2/\IZ_n$ by the singular Taub-NUT space and
 keep the radius of the Kaluza-Klein circles of the Taub-NUT space
  to be much larger than the 11-dimensional Plank length. Then T-dualize along two circles to get IIB string theory on a circle.  Under this duality, M2-branes turn into D3-branes
and the two Taub-NUT spaces turn into NS5/D5-branes.
Therefore, we obtain precisely the IIB brane configuration
discussed above, except that the $x^3$-direction is compactified. This duality transformation
seems to be a right step toward relating the M-theoretic aspect of our quiver theories to their type IIB interpretation.

\vspace{0.5cm}

\section*{Acknowledgments}
We thank Andreas Karch, Jun Nishimura and Piljin Yi for discussions.
Sm.L. is grateful to
the organizers of the INT workshop ``From Strings to Things'' at the University of Washington in Seattle
where part of the work was done.
K.M.L. and J.P. are  supported in part by the KOSEF SRC Program through CQUeST at
Sogang University. K.M.L. is supported in part by KRF Grant No. KRF-2005-070-C00030, and the KRF National
Scholar program.
Sm.L. is supported in part by the KOSEF Grant R01-2006-000-10965-0 and the
Korea Research Foundation Grant KRF-2007-331-C00073. J.P. is supported in part   by the Stanford Institute for
Theoretical Physics.

\newpage

\centerline{\large \bf Appendix}

\appendix

\section{Notations and Conventions}

\paragraph{Spinor calculus}
Spinor indices run $\alpha=+,-$. Indices are raised or lowered
by real antisymmetric matrices $\epsilon_{\alpha\beta}$ and
$\epsilon^{\alpha\beta}$ satisfying
$\epsilon_{\alpha\beta}\epsilon^{\beta\gamma}=\delta_\alpha^{~\gamma}$.
\[
 \psi_\alpha = \epsilon_{\alpha\beta}\psi^\beta,~~~~~
 \psi^\alpha = \epsilon^{\alpha\beta}\psi_\beta.
\]
Space-time metric has signature $(-++)$. 
The $\gamma$-matrices $(\gamma^\mu)_\alpha^{~\beta}$ satisfy the relations
\[
 \gamma^\mu\gamma^\nu+\gamma^\nu\gamma^\mu=2\eta^{\mu\nu},~~~~~
 \gamma^{[\mu}\gamma^\nu\gamma^{\rho]}=\varepsilon^{\mu\nu\rho}.~~~~~
 (\varepsilon^{012}=1)
\]
The matrices $(\epsilon\gamma^\mu)^{\alpha\beta}$ are real symmetric.
Vectors such as $x^\mu$ and $\partial_\mu$ are expressed as bi-spinors
\[
 x^{\alpha\beta}= x^\mu(\epsilon\gamma_\mu)^{\alpha\beta},~~~~
 \partial_{\alpha\beta}=-(\gamma^\mu\epsilon)_{\alpha\beta}\partial_\mu.
\]
Spinor indices in the standard position will be omitted.
\[
 \psi\theta\equiv\psi^\alpha\theta_\alpha,~~~~
 \theta^2=\theta^\alpha\theta_\alpha,~~~~
 \psi\gamma^\mu\theta=\psi^\alpha(\gamma^\mu)_\alpha^{~\beta}\theta_\beta,~~~~
 {\rm etc}.
\]
\paragraph{Superspace and superfields}
${\cal N}=1$ superspace coordinates are $x^{\alpha\beta}$ and
$\theta^\alpha$ (both real).
Supersymmetry algebra is realized in terms of supertranslations
\[
 P_{\alpha\beta}=i\partial_{\alpha\beta},~~~
 Q_\alpha=i\partial_\alpha+\theta^\beta\partial_{\beta\alpha}.
\]
Supercovariant derivatives are defined by
\begin{equation}
 D_{\alpha\beta}=\partial_{\alpha\beta},~~~~
 D_\alpha=\partial_\alpha+i\theta^\beta\partial_{\beta\alpha}.~~~~~
 \{D_\alpha,D_\beta\}=2iD_{\alpha\beta}.
\end{equation}
Components of general superfields $\Phi$ are defined by
$D_\alpha\cdots D_\beta\Phi|$, where $|$ takes the zeroth term
of the power series in $\theta$.
Supersymmetry acts on superfields as superderivatives.
Its action on components $(\cdots)|$ can be written as
\[
 \delta_\xi(\cdots)| ~\equiv~ -i \xi^\alpha(Q_\alpha\cdots)|.
\]
For example, scalar superfield $\Phi$ takes the form
\begin{equation}
 \Phi = \phi + i\theta\psi - \frac i2\theta^2F,~~~
 \phi = \Phi|,~~~
 \psi_\alpha= -iD_\alpha\Phi|,~~~
 F= -\frac i2D^2\Phi|,
\end{equation}
where $D^2\equiv D^\alpha D_\alpha$.
Supersymmetry transformation law of components becomes
\begin{equation}
 \delta\phi = i\xi^\alpha\psi_\alpha,~~~~
 \delta\psi_\alpha =
  -\partial_\alpha^{~\beta}\phi\xi_\beta -F\xi_\alpha,~~~~
 \delta F = i\xi^\beta\partial_\beta^{~\alpha}\psi_\alpha.
\end{equation}

Gauge symmetry is described by superconnections on the superspace.
As a simple example, consider a column vector $\Phi$ and a row vector
$\bar\Phi$ of scalar multiplets.
We wish to gauge the symmetry of unitary rotations
\[
 \Phi'=U\Phi,~~~~\bar\Phi'=\bar\Phi U^\dagger .
\]
To do this, we introduce the covariant derivatives
$\nabla_\alpha=D_\alpha+{\cal A}_\alpha$ and
$\nabla_\mu=D_\mu+{\cal A}_\mu$.
Gauge transformation acts on the superconnection in the standard manner,
\[
 {\cal A}'_\alpha = U{\cal A}_\alpha U^\dagger+UD_\alpha U^\dagger,~~~~
 {\cal A}'_\mu      = U{\cal A}_\mu      U^\dagger+UD_\mu      U^\dagger.
\]
The correct set of component fields is obtained by solving
the Bianchi identity that arise from
$\{\nabla_\alpha,\nabla_\beta\}=2i\nabla_{\alpha\beta}$.
The explicit solution reads
\begin{equation}
 [\nabla_\alpha,\nabla_{\beta\gamma}]=
 i\epsilon_{\alpha\beta}{\cal W}_\gamma+i\epsilon_{\alpha\gamma}{\cal W}_\beta,
\end{equation}
\begin{equation}
  \nabla^\alpha {\cal W}_\alpha=0,~~~~~
  \nabla_\alpha {\cal W}_\beta
 =\nabla_{(\alpha}{\cal W}_{\beta)}={\cal F}_{\alpha\beta}.
\end{equation}
Here ${\cal W}_\alpha$ is the gaugino superfield and ${\cal F}_{\alpha\beta}$
is the gauge field strength,
\begin{equation}
 {\cal F}_{\alpha\beta}=
 -\frac12{\cal F}_{\mu\nu}(\gamma^{\mu\nu}\epsilon)_{\alpha\beta},~~~~~
 {\cal F}_{\mu\nu}=\partial_\mu{\cal A}_\nu-\partial_\nu{\cal A}_\mu
 +[{\cal A}_\mu,{\cal A}_\nu].
\end{equation}
In Wess-Zumino gauge, these superfields take the form
\begin{eqnarray}
 {\cal A}_\alpha &=& i\theta^\beta A_{\alpha\beta}+\theta^2\chi_\alpha, \nn\\
 {\cal A}_{\alpha\beta} &=&
  A_{\alpha\beta} -i\theta_\alpha\chi_\beta -i\theta_\beta\chi_\alpha
 +\frac i2\theta^2F_{\alpha\beta}, \nn\\
 {\cal W}_\alpha &=& \chi_\alpha + \theta^\beta F_{\alpha\beta}
 -\frac i2\theta^2\nabla_\alpha^{~\beta}\chi_\beta.
\end{eqnarray}

\paragraph{General gauge theory}

Let us spell out the full Lagrangian for $\CN=1$ supersymmetric gauge theories.
First, the gauge kinetic term consists of the Chern-Simons
and Yang-Mills terms
\begin{eqnarray}
 \frac{k}{16\pi}\int d^2\theta{\rm Tr}
 \left(-iA{\cal W}
       +{\ts\frac16}\{A^\beta,A^\gamma\}A_{\beta\gamma}\right)
 &=& \frac{k}{4\pi}{\rm Tr}\left[
   \varepsilon^{\mu\nu\rho}(A\partial A+{\ts\frac23}A^3)_{\mu\nu\rho}
   +i\chi\chi\right],
\nn \\
  -\frac{1}{8g^2}\int d^2\theta{\rm Tr}{\cal W}^2
 &=& \frac{1}{4g^2}{\rm Tr}\left(F^{\mu\nu}F_{\mu\nu}-2i\chi\sla\nabla\chi\right).
\end{eqnarray}
The matter kinetic term and the superpotential term are given by
\begin{eqnarray}
  \frac14\int d^2\theta\nabla\bar\Phi\nabla\Phi
 &=& -\nabla^\mu\bar\phi\nabla_\mu\phi+\bar FF
     +i\bar\psi\sla\nabla\psi
     -i\bar\psi\chi\phi+i\bar\phi\chi\psi,
 \nn\\
  \frac i2\int d^2\theta W(\Phi) &=& -\frac i2W_{ij}\psi^i\psi^j-W_iF^i.
\end{eqnarray}
Here the matter superfields are taken to be real in the second line.
We have also slightly redefined the matter component fields,
\begin{equation}
 \phi=\Phi|,~~~~
 \psi_\alpha= -i\nabla_\alpha\Phi|,~~~~
 F= -\frac i2\nabla^\alpha\nabla_\alpha\Phi|.
\end{equation}
The supersymmetry transformation rules for the component fields are
\begin{equation}
 \delta\phi= i\xi\psi,~~~~
 \delta\psi= -\sla\nabla\phi\xi-F\xi,~~~~
 \delta F  = i\xi\sla\nabla\psi-i\xi\chi\phi.
\end{equation}
\begin{equation}
 \delta_\xi A_\mu = -i\xi\gamma_\mu\chi,~~~~
 \delta_\xi\chi = -\frac12F_{\mu\nu}\gamma^{\mu\nu}\xi.
\end{equation}

\end{document}